\documentclass[12pt,twocolumn]{aastex6}
\usepackage{amssymb,amsmath}
\usepackage{hyperref}
\usepackage{ulem}
\usepackage{color}
\usepackage{natbib}
\usepackage{fancyhdr}

\slugcomment{}

\shorttitle{\lowercase{h}COSMOS}
\shortauthors{Damjanov et al.}

\begin{document}

\title{\lowercase{h}COSMOS: a  dense spectroscopic survey of $\lowercase{r}\leqslant21.3$ galaxies in the COSMOS field}

\author{Ivana Damjanov\altaffilmark{1, 2}, H. Jabran Zahid\altaffilmark{3}, Margaret J. Geller\altaffilmark{3},  Daniel G. Fabricant\altaffilmark{3}, Ho Seong Hwang\altaffilmark{4}}
\altaffiltext{1}{Harvard-Smithsonian Center for Astrophysics, 60 Garden Street, Cambridge, MA 02138, USA; \href{mailto:idamjanov@cfa.harvard.edu}{idamjanov@cfa.harvard.edu}}
\altaffiltext{2}{Department of Astronomy and Physics, Saint Mary's University, 923 Robie Street, Halifax, NS B3H 3C3, Canada; \href{mailto:Ivana.Damjanov@smu.ca}{Ivana.Damjanov@smu.ca}}
\altaffiltext{3}{Smithsonian Astrophysical Observatory, 60 Garden Street, Cambridge, MA 02138, USA}
\altaffiltext{4}{Quantum Universe Center, Korea Institute for Advanced Study, 85 Hoegiro, Dongdaemun-gu, Seoul 02455, Republic of Korea}

\begin{abstract}
We describe the hCOSMOS redshift survey of the COSMOS field conducted with the Hectospec spectrograph on the MMT. In the central 1~deg$^2$, the hCOS20.6 subset of
the survey is  $>90\%$ complete to a limiting $r=20.6$. The hCOSMOS survey includes 1701 new redshifts in the COSMOS field. We also use the total of 4362 new and remeasured objects to derive the age sensitive D$_n4000$ index over the entire redshift interval $0.001\lesssim z\lesssim0.6$. For $85\%$ of the quiescent galaxies in hCOS20.6, we measure the central line-of-sight velocity dispersion. To explore potential uses of this survey, we combine previously measured galaxy sizes, profiles and stellar masses with the spectroscopy. The comparison reveals the known relations among structural, kinematic, and stellar population properties. We also compare redshift and D$_n4000$ distributions of hCOS20.6 galaxies with SHELS; a complete spectroscopic survey of 4~deg$^2$ observed to the same depth. The redshift distributions in  the two fields are very different but the D$_n4000$ distribution is remarkably similar. The relation between velocity dispersion and stellar mass for massive hCOS20.6 galaxies is consistent with the local relation from SDSS. Using measured velocity dispersions, we test a photometric proxy calibrated to galaxies in the local universe.  The systematic differences between the measured and photometric proxy velocity dispersions are correlated with galaxy dynamical and stellar population properties highlighting the importance of direct spectroscopic measurements. 

\end{abstract}


\section{Introduction}

Dense redshift surveys of large volumes are powerful tools for studying galaxy cosmic evolution and its relation to the underlying matter distribution. Large area surveys have transformed our view of the distribution of galaxies in the nearby universe \citep[e.g.,][]{Geller1989, Loveday1992, Shectman1996, Vettolani1997, daCosta1998, York2000, Colless2001, Jones2009, Baldry2010}. At high redshift ($z\gtrsim1$) spectroscopic studies of extended areas are currently not feasible. Most surveys at higher redshift are confined to smaller areas ($\lesssim2$~deg$^2$) (e.g, \citealt{Davis2003, Steidel2003, Wirth2004, Noll2004, Abraham2004, LeFevre2005, Newman2013, LeFevre2015}). The VIPERS survey (\citealt{Guzzo2014}) covers 24~deg$^2$. At intermediate redshifts, dense, complete surveys over areas comparable to VIPERS are feasible; e.g., at  $0.1<z<0.6$ the SHELS survey covers two widely separated 4~deg$^2$ fields \citep{Geller2014, Geller2016}. The sparser AGES survey focuses on AGN and covers a 7.7~deg$^2$ area to similar depth \citep{Kochanek2012}.

High-resolution imaging adds another dimension to spectroscopic surveys targeting the intermediate-redshift universe. The combination of photometric and spectroscopic data enables evolutionary studies of relations among the spectroscopic and structural properties of galaxies. The largest extragalactic field surveyed with the Hubble Space Telescope (HST), the $\sim2$deg$^2$ COSMOS field, has thus been the preferred target for a suite of intermediate- and high-redshift spectroscopic campaigns (e.g, zCOSMOS, \citealt{Lilly2007, Lilly2009}; PRIMUS, \citealt{Coil2011}; VUDS, \citealt{LeFevre2015}; FMOS-COSMOS, \citealt{Silverman2015}; LEGA-C; \citealt{vanderWel2016}). At $0.1<z<0.6$, broad spectroscopic wavelength coverage and a high level of completeness are critically important for accurately tracing the relations between photometric and spectroscopic properties of galaxies.    

Here we describe hCOSMOS, a dense redshift survey of the COSMOS field conducted with the MMT Hectospec multi-fiber spectrograph.  A subset of hCOSMOS, hCOS20.6, covers the central $\sim1$~deg$^2$
of the field. hCOS20.6 includes $>90\%$ of galaxies with $r-$band magnitude $r\leq20.6$ (i.e., hCOS20.6 is 90\% complete to the limiting magnitude r=20.6). Our straightforward survey design along with  the Hectospec spectra covering the range $3700-9100$~\AA\ enable explorations of the interplay between the dynamical, structural, and stellar population properties of galaxies and their environments at intermediate redshifts. 

We used some of these data to study the properties of the quiescent galaxy population at $0.1<z<0.6$ \citep{Damjanov2015b, Zahid2016a}. Taking advantage of available high-resolution imaging, we select a sample of massive compact quiescent galaxies and explore the dependence of their properties on the surrounding galaxy density field \citep{Damjanov2015b}. Massive compact quiescent galaxies lie preferentially in denser regions at intermediate redshifts. This trend is driven mainly by the large stellar masses of these objects; compact galaxies tend to be massive and massive galaxies are preferentially located in dense regions.  

We explored the dynamical properties of quiescent galaxies in hCOSMOS by examining the stellar mass fundamental plane at $0.1<z<0.6$ \citep{Zahid2016a}. The orientation of the stellar mass fundamental plane is independent of redshift for these systems and the zero-point appears to evolve by a small amount $\sim0.04$~dex. Compact quiescent galaxies fall on the same relation as the extended objects. The hCOSMOS dataset confirms that the compact quiescent population does not constitute a special class of objects; the compact population is  the tail of the size and stellar mass distribution of the general quiescent population \citep[see also][]{Zahid2015}. 

Here we provide the hCOSMOS spetro-photometric dataset. To demonstrate the quality of the survey data, we provide a few applications of the data. We recover the known relations among structural, kinematic, and stellar population properties of the quiescent population. We demonstrate that the predicted and measured velocity dispersion differ systematically potentially producing spurious effects. These differences highlight the importance of direct spectroscopic measurements.
 
We describe hCOSMOS and hCOS20.6  in Section~\ref{data}. We examine the completeness of hCOS20.6 in Section~\ref{com}. We explore combined spectroscopic and photometric properties of the hCOS20.6 galaxies in Section~\ref{properties} and compare salient characteristics of the hCOS20.6 sample with the larger area SHELS survey to similar depth.  We test the photometric proxy for inferring velocity dispersion in Section~\ref{proxy}. We conclude in Section~\ref{conclude}. We adopt the standard cosmology $(H_0,\, \Omega_m,\, \Omega_ \Lambda) = (70$~km~s$^{-1}$~Mpc$^{-1},\, 0.3,\, 0.7)$ and AB magnitudes throughout.

\section{Dataset}\label{data}

\subsection{Photometry}\label{photo}
\subsubsection{Target Selection}\label{targets}

The UltraVISTA photometric catalog\footnote{\url{http://vizier.cfa.harvard.edu/viz-bin/VizieR-3?-source=J/ApJS/206/8/catalog}}\citep{Muzzin2013} provides the basis for the spectrosocpic survey. The UltraVISTA catalog covers 1.6~deg$^2$ of the COSMOS field and includes point-spread function (PSF) matched photometry in 30 photometric bands over the $0.15-24\, \mu\mathrm{m}$ wavelength range \citep{Martin2005, Taniguchi2007, Capak2007, Sanders2007, McCracken2012}. We target galaxies with $r-$band magnitudes $17.77<r<21.3$ for spectroscopy. The bright limit of r = 17.77 is the limiting magnitude of the Sloan Digital Sky Survey (SDSS) main galaxy sample \citep{Strauss2002}.

We  selected extended objects in HST images \citep{Scarlata2007, Sargent2007}. There are 10750 $17.77<r<21.3$ UltraVISTA targets within the HST-ACS footprint of the COSMOS field. A large fraction of these systems (6709 objects or $62.5\%$) are classified as galaxies based on the STELLARITY flag ($=0$) in the Zurich Structure \& Morphology Catalog\footnote{\url{http://irsa.ipac.caltech.edu/data/COSMOS/tables/morphology/cosmos_morph_zurich_1.0.tbl}}. These 6709 COSMOS galaxies are the Hectospec targets (Section~\ref{obs}). 

We conducted observations in the Spring of 2015 and 2016. For the first observing run we targeted red galaxies ($g-r>0.8$; $r-i>0.2$) without publicly available secure redshifts from zCOSMOS \citep{Lilly2007, Lilly2009} and/or SDSS DR12 \citep{Alam2015}. During the 2016 observing run (Section~\ref{obs}) we targeted all galaxies with $17.77<r<21.3$ that we did not observe previously with Hectospec.  We also re-observed 2661 galaxies observed by zCOSMOS  to take advantage of the broader Hectospec wavelength coverage. Our final survey, hCOSMOS,  has no color selection. The spectra enable measurement of the D$_n4000$ index for all objects with spectroscopy (Section~\ref{dn4000}).

\subsubsection{Structural Parameters}\label{structure}

In addition to the extended COSMOS source selection,  we also check the structural parameters reported in the Zurich Structure \& Morphology Catalog \citep{Sargent2007} measured using Galaxy IMage 2D (GIM2D) software \citep{Simard2002}. GIM2D fits a two-dimensional surface brightness model convolved with the PSF to galaxy images. \citet{Sargent2007} measure the half-light radius $R_e$,
the S\'ersic index $n$ and the ellipticity $e$ by fitting a single \citet{Sersic1968} model. In Sections~\ref{size-mass}~and~\ref{proxy} we use the circularized half-light radius $R_{e,c}=R_e\times\sqrt{(1-e)}$ and S\'ersic index $n$.

\subsubsection{Stellar Masses}\label{mass}

We derive stellar masses using publicly available $ugriz-$band photometry of the COSMOS field \citep{Capak2007, Muzzin2013}. We estimate the mass-to-light (M/L) ratio for each galaxy by $\chi^2$ fitting the observed photometry with  synthetic spectral energy distributions (SEDs). The SED shape constrains the stellar M/L ratio used to convert luminosity to stellar mass. We fit the observed photometry with {\sc Lephare}\footnote{\url{http://www.cfht.hawaii.edu/$\sim$arnouts/LEPHARE/cfht\_lephare/ lephare.html}} \citep{Arnouts1999, Ilbert2006} using the stellar population synthesis models of \citet*{Bruzual2003} and the \citet{Chabrier2003} initial mass function (IMF). The stellar population models have three metallicities ($Z = 0.004$, 0.008 and 0.02) and exponentially declining star formation histories (star formation rate $\propto e^{-t/\tau}$) with e-folding times of $\tau = 0.1,0.3,1,2,3,5,10,15$ and $30$ Gyr. Synthetic SEDs are generated from these models by varying the extinction and stellar population age. Extinction is applied to the synthetic SEDs by adopting the \citet{Calzetti2000} extinction law and by allowing $E(B-V)$ to range from 0 to 0.6. The stellar population ages range between 0.01 and 13 Gyr. For each set of parameters, the procedure yields a distribution for the best-fit M/L ratio and stellar mass. We adopt the median of the stellar mass distribution as our estimate.

\subsection{Spectroscopy}\label{spec}

\subsubsection{Observations}\label{obs}

We observed hCOSMOS galaxies with Hectospec mounted on the 6.5~m MMT \citep{Fabricant1998, Fabricant2005}. Hectospec is a 300-fiber optical spectrograph with an $\sim1$ square degree field of view (FOV). The instrument FOV is well matched to the size of the COSMOS field. To maximize the completeness of hCOSMOS we targeted field positions around the center (R.A.$_\mathrm{2000}=10^\mathrm{h}00^\mathrm{m}29^\mathrm{s}$, Dec.$_\mathrm{2000}=+02\arcdeg12\arcmin21\arcsec$) of the COSMOS field and prioritized galaxies according to their $r-$band brightness, filling empty fibers with $r>21.3$ targets. Fiber positions were optimized using software developed by \citet{Roll1998}. 

 The Hectospec spectra cover the wavelength range $3700-9100$~\AA\ at a resolution of $R\sim1500$. At our faint limiting magnitude of $r=21.3$ an integration time of $1$~h yields a redshift in optimal observing conditions ($\lesssim1\arcsec$ seeing, dark time, airmass~$\sim1$). We conducted observations during dark and gray nights in February 2015, April 2015, February 2016, and March 2016, with $\sim1\arcsec$ seeing (Hectospec fiber diameter is $1.5 \arcsec$).  For brighter targets with $r<21.3$, typical observing conditions and $1$~h exposure yield a redshift (Section~\ref{redshift}), a D$_n4000$ index (Section~\ref{dn4000}), and a central velocity dispersion (for $\sim 80$\% of the objects; Section~\ref{veldisp}). We obtained 5492 science quality spectra (out of which 1405 are duplicates) in varying conditions. We include additional 275 spectra of galaxies in the COSMOS field from the Hectospec data archive. Thus, the hCOSMOS sample includes 4362 unique galaxies. Among these objects, 2661 have a redshift in the zCOSMOS catalog and 1701 redshifts are completely new. For the galaxies that overlap with zCOSMOS, the Hectospec spectra yield a redshift with a smaller error along with broader wavelength coverage. The broader spectral range of Hectospec allows for the D$_n4000$ index measurements across the redshift interval of the sample ($0.001<z\lesssim0.7$; Section~\ref{dn4000}). The spectral resolution enables velocity dispersion measurements for quiescent galaxies with high-quality spectra (Section~\ref{veldisp}).
 
\begin{deluxetable}{lcc}
\tabletypesize{\small}
\tablecaption{Properties of the hCOSMOS Survey\label{table1}}
\tablewidth{7in}
\tablehead{
\colhead{Property} & \multicolumn{2}{c}{Value}\\
\colhead{} & \colhead{hCOSMOS} & \colhead{hCOS20.6}
}
\colnumbers
\startdata 
{$r-$band limit [mag]} & 21.3 & 20.6\\
{Area [deg$^2$]}  & 1.59 & 0.89\\
$N_\mathrm{photo}$ & 6709 & 2041\\  
$N_\mathrm{spec}$ & 4362 & 1968\\
$z$ & (0.001, 0.7) & (0.001, 0.55)\\
$\widetilde{z}$ & 0.32 & 0.27\\
$N_\mathrm{SDSS}$\tablenotemark{a} &  ... & 37\\
$N_\mathrm{hCOS\,\&\,zCOS}$\tablenotemark{b} & 2661 & 1283 \\
$N_{\mathrm{D}_n4000}$ & 4341 & 1968\\
$N_{\mathrm{D}_n4000,\,\mathrm{hCOS\,\&\,zCOS}}$\tablenotemark{c}  & 383 & 75\\
$N_{{\mathrm{D}_n4000}>1.5}$ & 1713 & 901\\
$N_{{\mathrm{D}_n4000}>1.5,\, {\sigma_0}}$\tablenotemark{d} & 762  & 762\\
\enddata
\tablecomments{
\tablenotetext{a}{Number of SDSS spectra added to the hCOS20.6 sample (Section~\ref{com})}
\tablenotetext{b}{Number of overlaps with zCOSMOS 20K spectroscopic sample \citep{Lilly2007, Lilly2009}}
\tablenotetext{c}{Number of overlaps with zCOSMOS at redshift $z\geqslant0.44$ where D$_n4000$ index can be measured from zCOSMOS spectra}
\tablenotetext{d}{Number of velocity dispersion measurements that we report only for the hCOS20.6 quiescent sample (see Section~\ref{veldisp} for details)}
}
\end{deluxetable}

\subsubsection{Redshifts}\label{redshift}

\begin{figure*}
\begin{centering}
\includegraphics[scale=0.65]{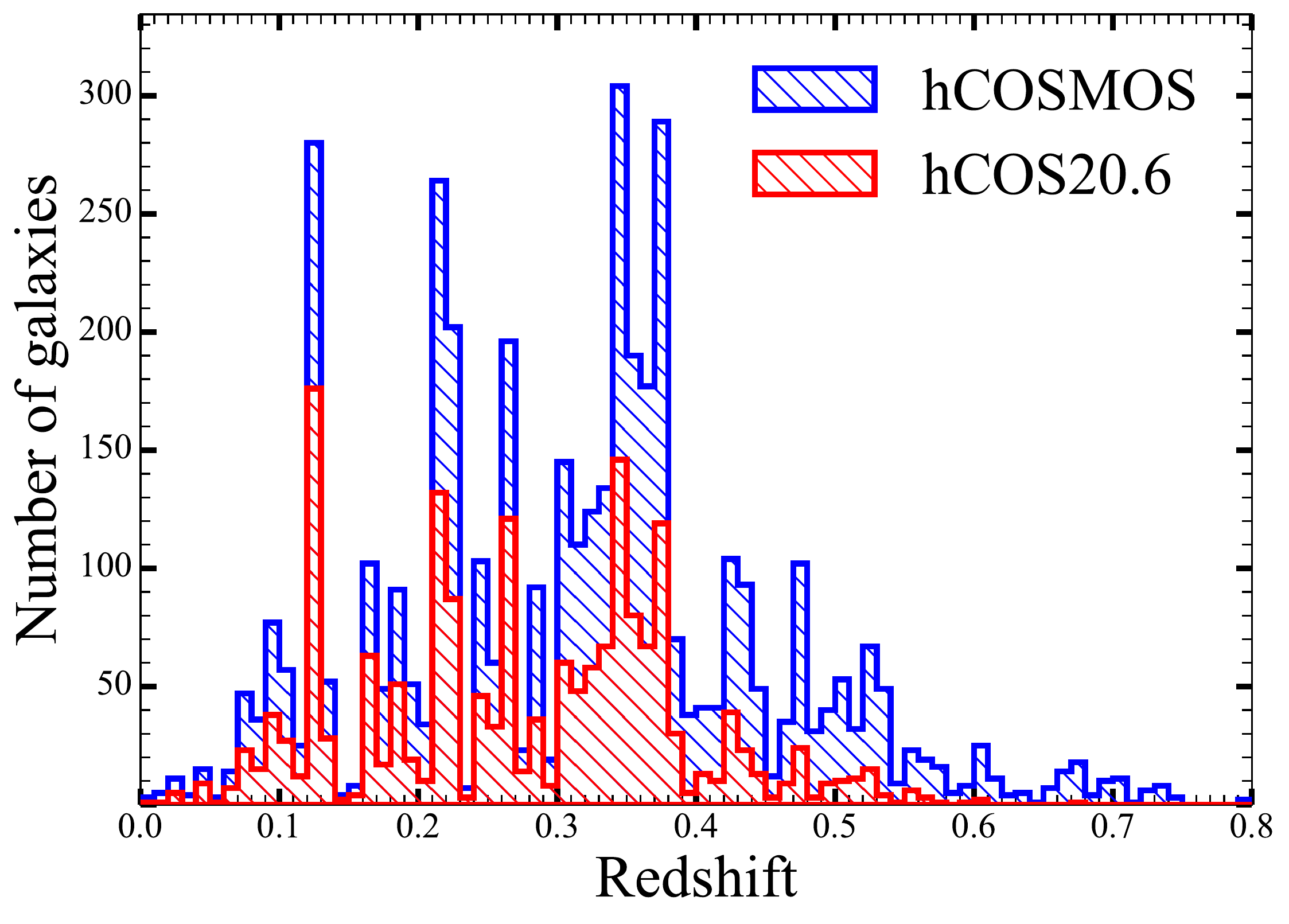}
\caption{Redshift distribution of hCOSMOS galaxies ($r<21.3$) and of the magnitude limited hCOS20.6 sample ($r<20.6$; Section~\ref{com}). The same  density peaks are apparent in the redshift
range  $0.001<z<0.4$.   
\label{f1}}  
\end{centering}
\end{figure*}

\begin{figure*}
\begin{centering}
\hspace*{-0.35in}
\includegraphics[scale=0.4]{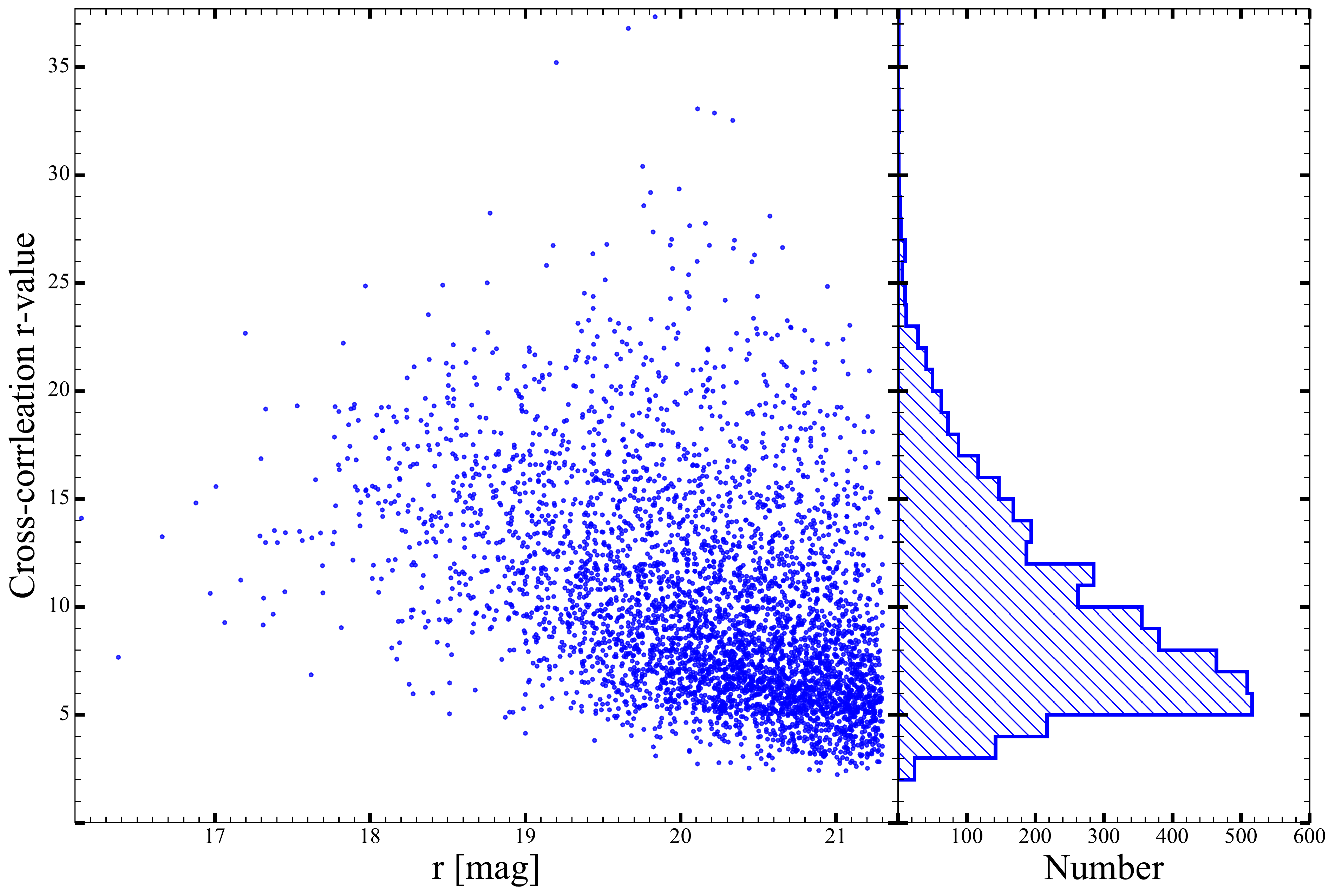}
\caption{Left: cross-correlation r-value \citep{Tonry1979} 
as a function of $r-$band magnitude for 4362 hCOSMOS galaxies with a reliable redshift (Table~\ref{table1}). Right: distribution of the cross-correlation $r-$value for hCOSMOS spectra. We test the low $r-$value redshift measurements by cross-matching the sample with the zCOSMOS-bright 20K catalog. \label{f2}}  
\end{centering}
\end{figure*}

\begin{figure}
\begin{centering}
\includegraphics[scale=0.25]{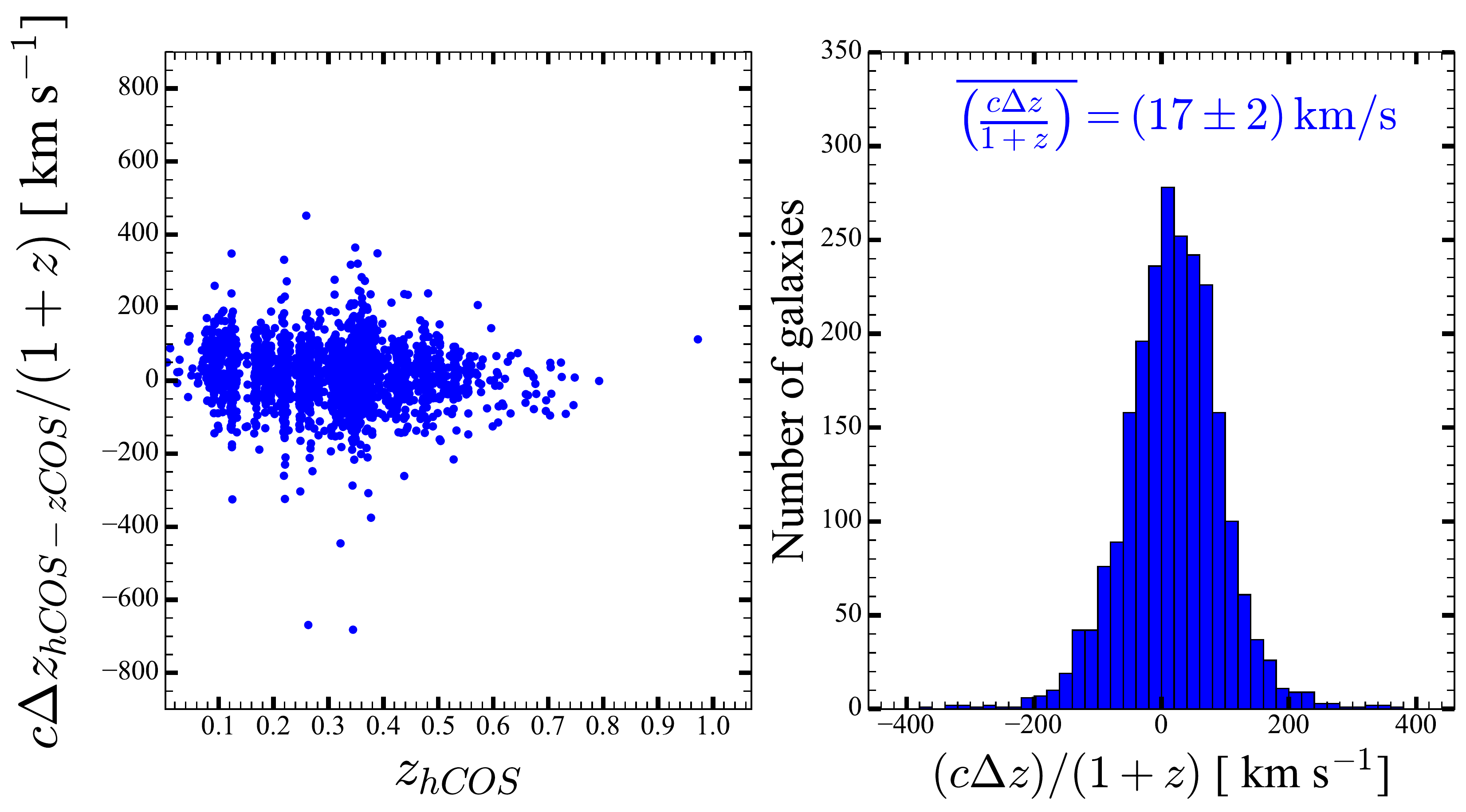}
\caption{Normalized difference between hCOSMOS and zCOSMOS redshifts as a function of the hCOSMOS  spectroscopic redshift. We use hCOSMOS galaxies with the cross-correlation r-value$>5$ for this comparison. Distribution of $\Delta z$ normalized by $(1+z)$ (right) for the measurements in the left-hand panels. Labels indicate the mean and standard deviation. \label{f3}}  
\end{centering}
\end{figure}

Hectospec data are reduced with HSRED v2.0, developed by the Telescope Data Center. This pipeline is a revision of the original IDL pipeline  by Richard Cool\footnote{\url{http://www.mmto.org/book/export/html/55}}. The pipeline provides one-dimensional, wavelength calibrated, and sky subtracted spectra. We derive a relative flux calibration using an average correction for the relative throughput of the detector as a function of wavelength. This correction is based on repeat observations of standard stars. The relative throughput between 4000 and 8000~\AA\ remains remarkably stable over years \citep[see Figure~2 of][]{Fabricant2008}. 

Redshifts are measured by cross correlating observed spectra against a library of template spectra \citep{Kurtz1998}. The pipeline returns the $r-$value, a relative amplitude of the cross-correlation peak \citep{Tonry1979}, as a measure of the quality of the redshift. Based on visual inspection, we conclude that redshifts with $r-$values $>5$ are secure. To maximize completeness, we cross-match redshifts with $r-$values $<5$ with the highest quality redshifts (i.e., redshifts with a spectroscopic verification rate $>99\%$)\footnote{\url{http://www.eso.org/sci/observing/phase3/data_releases/zcosmos_dr3_b2.pdf}} of the 20K zCOSMOS-bright catalog \citep{Lilly2007, Lilly2009}. We consider redshift measurements with $r-$values~$<5$ reliable if they are within $\pm2000$~km~s$^{-1}$ of the corresponding zCOSMOS redshifts. Our final hCOSMOS sample includes 4362 distinct galaxies with reliable redshifts. Table~\ref{table1} provides an overview of the hCOSMOS survey and Table~\ref{table2} lists the spectroscopic measurements (redshift, D$_n$4000, and central velocity dispersion) and stellar mass estimates (Section~\ref{mass}) for the hCOSMOS sample. 

Figure~\ref{f1} shows the distribution of reliable hCOSMOS redshifts (blue histogram). The redshift distribution for $r<21.3$ hCOSMOS systems shows prominent overdensities at $z\sim0.12, 0.22, 0.26, 0.35, 0.37$. The same set of density peaks is evident in the distribution of redshifts for a complete magnitude limited subsample hCOS20.6 covering the central $\sim1$~deg$^2$ of the COSMOS field (red histogram, Section~\ref{com}).

Figure~\ref{f2} shows the $r-$value as a function of redshift for hCOSMOS galaxies (left-side panel) and the overall distribution of $r-$values for the sample. As expected, the cross-correlation value decreases with increasing $r-$band magnitude. Some systems ($\sim9\%$) in our sample have $r-$values $\lesssim5$. These objects have prominent spectral features in otherwise noisy spectra. As noted above, we include redshifts derived from low $r-$value spectra only if they are consistent with the values reported in the zCOSMOS catalog.

We use repeat observations to estimate the mean internal error in the Hectospec redshift. We limit the comparison to observations with $r-$values $\geqslant5$ and $z<0.7$. We estimate an error of $42$ and 26~km~s$^{-1}$ for absorption line and emission line systems, respectively. Our internal error estimates are consistent with estimates based on much larger Hectospec datasets \citep[SHELS,][]{Geller2014,Geller2016}. 

We compare the Hectospec redshift measurements with  the  zCOSMOS-bright (20K)  sample. There are 2661 repeat observations (Table~\ref{table1}) and the offset between hCOSMOS and zCOSMOS redshift measurements is $17\pm2$~km~s$^{-1}$ (Figure~\ref{f3}); this is smaller than the typical statistical uncertainties in an individual measurement. The root-mean-square scatter of 96~km~s$^{-1}$ is slightly lower than the typical redshift errors  added in quadrature for hCOSMOS and zCOSMOS ($110$~km~s$^{-1}$ is the typical zCOSMOS redshift error quoted by \citealt{Lilly2009}).

\subsubsection{Spectroscopic measurements: D$_n4000$ index}\label{dn4000}

The D$_n4000$ index is defined as the ratio of flux measured between $3850-3950$~\AA\ and
 $4000-4100$~\AA. These ranges are narrow to minimize the effects of reddening \citep{Balogh1999}. 
This spectral index measures the strength of the $4000$\AA\ break produced by a large number of absorption lines where ionized metals are the main contributors to the opacity. In young, hot stars the elements are multiply ionized, decreasing the line opacities. Thus the strength of the $4000$~\AA\ break is smaller for systems dominated by young stellar populations and it increases with the stellar population age. In Section~\ref{dn4000_sfq} we employ the magnitude limited subsample of hCOSMOS galaxies to demonstrate how the D$_n4000$ index discriminates between star-forming and quiescent galaxy populations \citep[e.g.,][]{Kauffmann2003, Woods2010, Geller2014}.

We measure D$_n4000$ for 4341 hCOSMOS galaxies (99.5\% of the targets with reliable spectroscopic redshifts, Table~\ref{table1}).  The only other spectroscopic survey of the COSMOS field with a comparable completeness level, zCOSMOS-bright, provides D$_n4000$ measurements only for galaxies with redshifts $z\gtrsim0.44$. This high redshift limit results from the narrower spectral range of the VLT VIMOS grism \citep[$5500-9650$~\AA,][]{Lilly2007}. In hCOSMOS galaxies without D$_n4000$ measurements are either a) faint, with $r-$band magnitudes close to the $r=21.3$~mag limit (the median $r-$band magnitude for these galaxies is  $\widetilde r=20.9$), and/or b) have a redshift where the rest-frame wavelength range used for D$_n4000$ measurements is contaminated by  strong night sky emission lines. Based on repeat observations of 1027 hCOSMOS objects, the typical error in D$_n4000$ is $0.057\, \times$~the measured value \citep[see also][]{Fabricant2008}.

\subsubsection{Spectroscopic measurements: velocity dispersion}\label{veldisp}

\begin{figure*}
\begin{centering}
\includegraphics[scale=0.75]{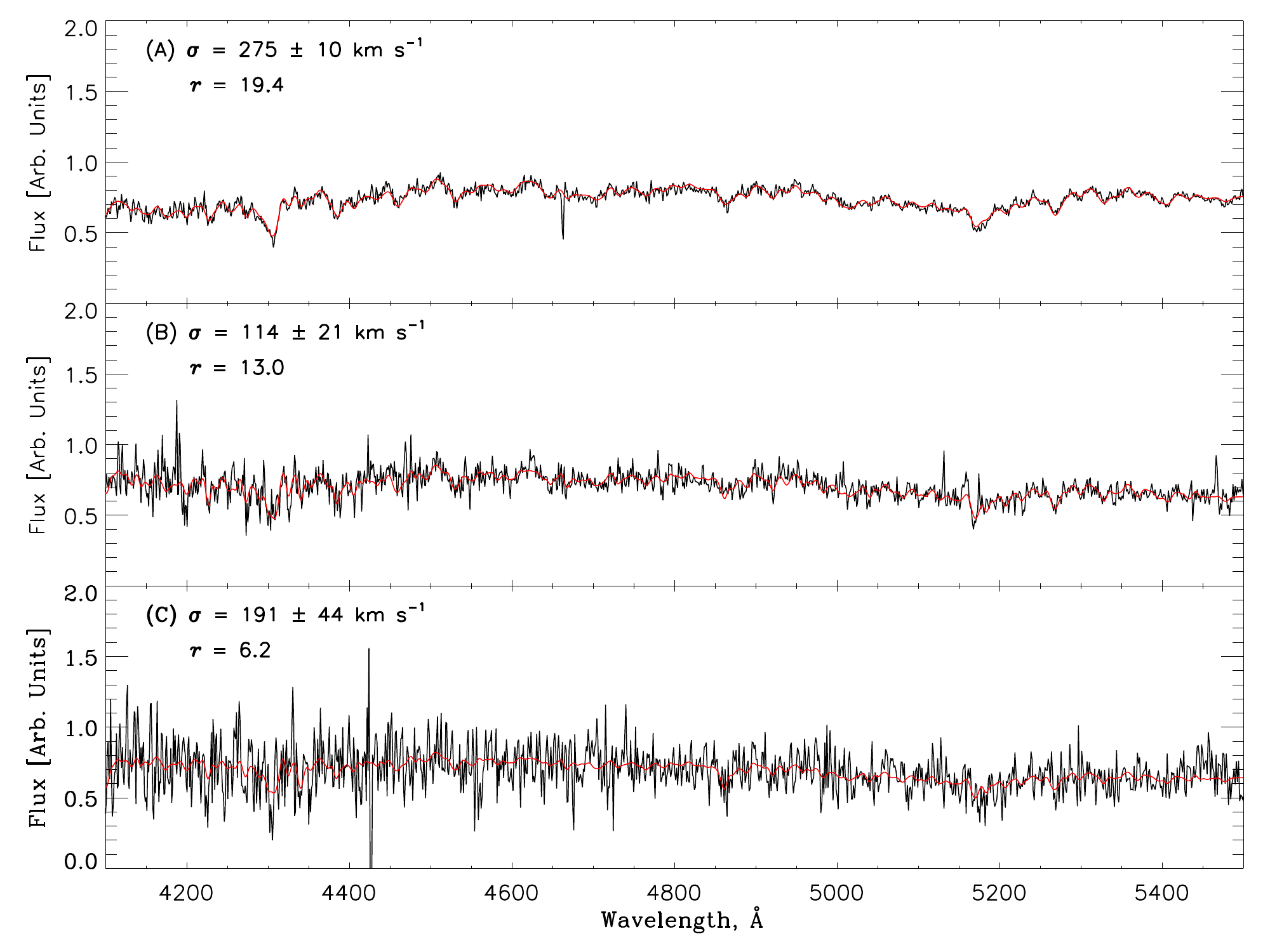}
\caption{Spectra of three quiescent galaxies in the spectral range used to determine velocity dispersion. The spectra range in data quality (described by $r-$value as in Section~\ref{redshift}) and velocity dispersion which are noted in each panel. The red curve in each plot is the best fit model. The data are smoothed by 5 pixels for display purposes. 
 \label{f4}}  
\end{centering}
\end{figure*}

We measure the central velocity dispersion using the University of Lyon Spectroscopic analysis Software\footnote{\url{http://ulyss.univ-lyon1.fr/}} \citep[ULySS;][]{Koleva2009}. We fit the spectrum with single age stellar population models calculated with PEGASE-HR code \citep{LeBorgne2004} and MILES stellar library \citep{SanchezBlazquez2006}. The fit is limited to $4100-5500$ $\mathrm{\AA}$ in the rest-frame. This wavelength range minimizes velocity dispersion errors and provides the most stable results \citep{Fabricant2013}. This spectral range is accessible with Hectospec out to $z\sim0.65$, the redshift range spanned by the hCOSMOS data. Models are convolved to the wavelength dependent spectral resolution of the Hectospec data taking  the line spread function into account \citep{Fabricant2013}. Models are parameterized by age and metallicity and are convolved with the line-of-sight velocity dispersion. The best-fit age, metallicity and velocity dispersion are determined by a $\chi^{2}$ minimization. Given the S/N of our observations and the resolution of Hectospec, velocity dispersions are typically reliable down to $\sim90$ km s$^{-1}$. Details of the velocity dispersions measurements with Hectospec along with systematic issues are discussed in \citet{Fabricant2013}.

Figure~\ref{f4} shows  spectra for three quiescent galaxies ranging in data quality and velocity dispersion. From a larger data set of $\sim2500$ repeat observations, we show that the velocity dispersions of quiescent galaxies agree within the observational uncertainties of the fits as shown in Figure ~\ref{f4} (Geller et al., in prep). From these same set of repeat observations, we determine that velocity dispersions measured for emission line galaxies are unreliable. Our sample selection criterion of $D_{n}4000>1.5$ (see Section~\ref{dn4000_sfq}) effectively removes emission line galaxies from the sample \citep{Woods2010}. We report 762 velocity dispersions measurements with $\chi^{2}<2$ for quiescent galaxies in the complete hCOS20.6 sample (Section~\ref{com}). 

\section{Completeness}\label{com}

\begin{figure*}
\begin{centering}
\includegraphics[scale=0.25]{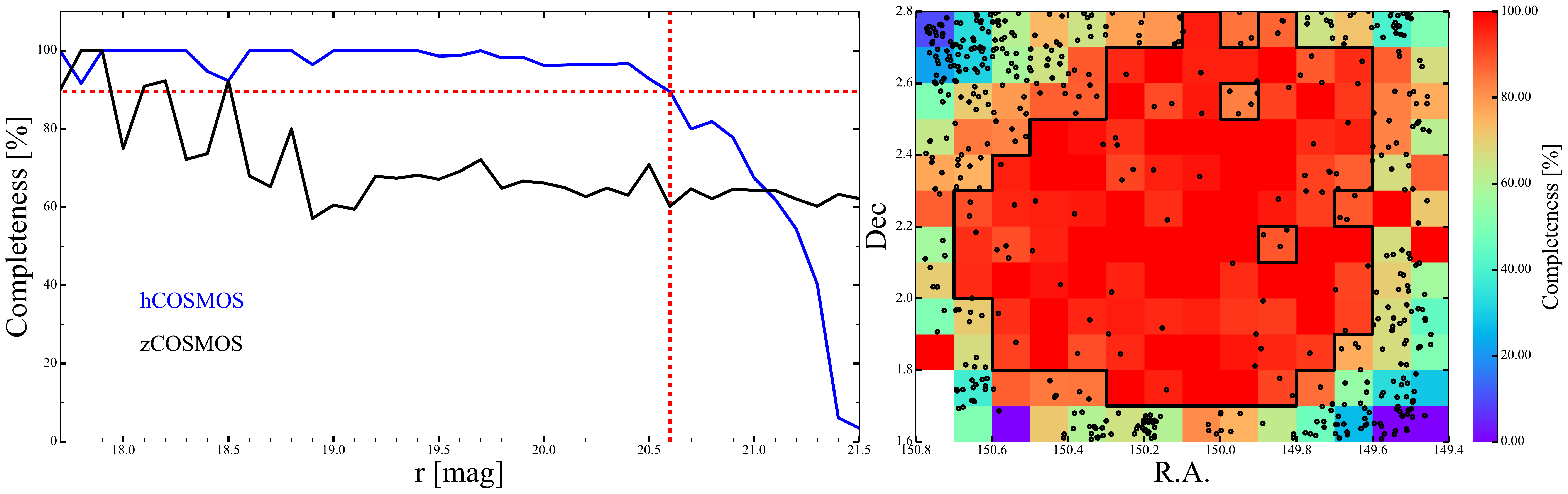}
\caption{Completeness of the hCOSMOS survey. The left sub-panel shows the differential completeness as a function of $r-$band magnitude (blue line) in comparison with the differential completeness of zCOSMOS (black line). The vertical red line at $r = 20.6$ marks the 90\% differential completeness limit for hCOSMOS in the area outlined by the black line in the lower sub-panel. The right sub-panel shows the spectroscopic completeness in $6\arcmin\times6\arcmin$ bins for hCOSMOS galaxies with $r<20.6$. The black line outlines the central 0.89~deg$^2$ covered by a subset of hCOSMOS galaxies, hCOS20.6; the differential spectroscopic completeness is $>90\%$ for galaxies with $r<20.6$ in this region. Black points indicate targets from the photometric UltraVISTA sample \citep{Muzzin2013} at $r<20.6$ and without a reliable hCOSMOS redshift.
\label{f5}}
\end{centering}
\end{figure*}

\begin{figure}
\begin{centering}
\includegraphics[scale=0.45]{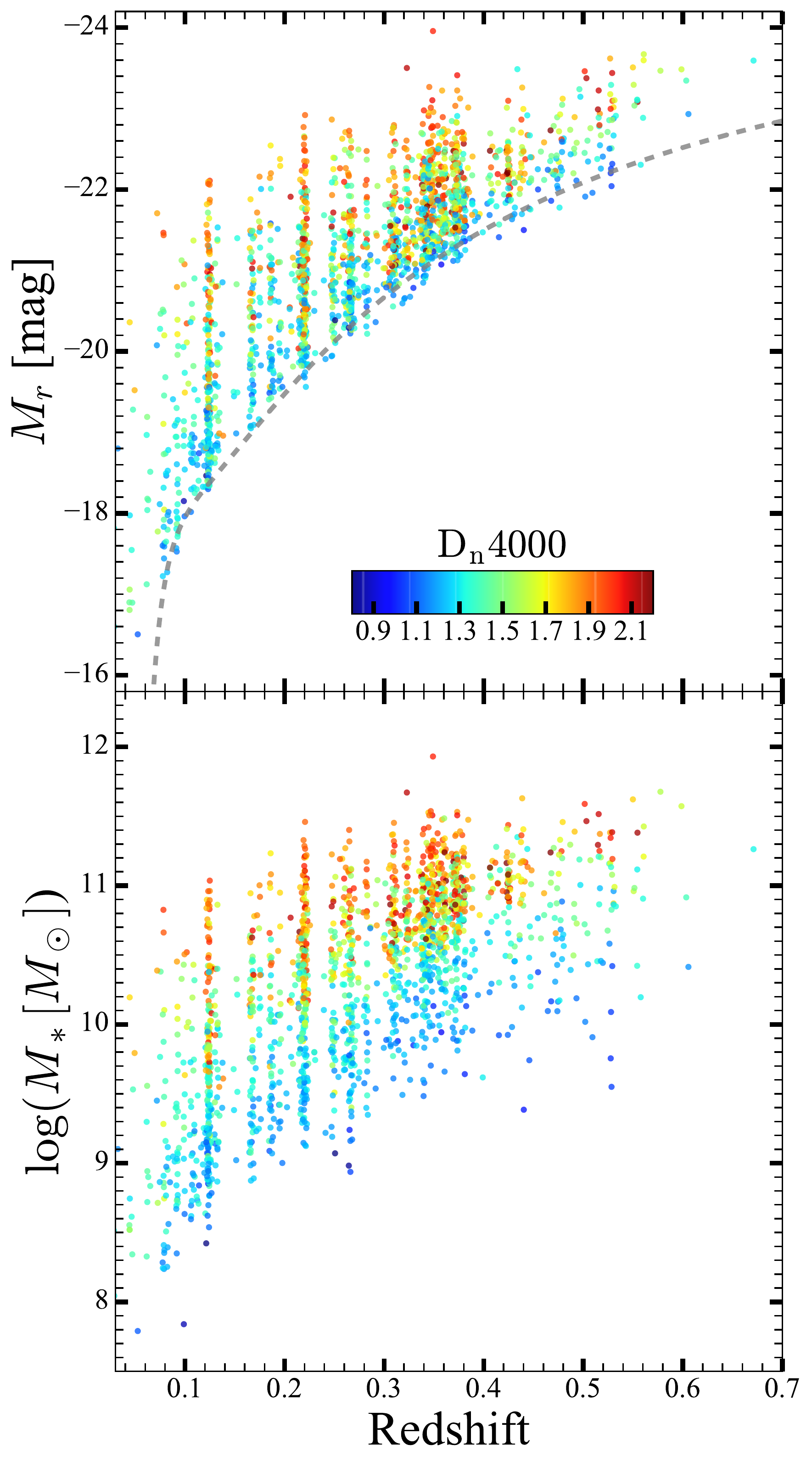}
\caption{Absolute r-band magnitude versus redshift (top) and stellar mass versus redshift (bottom) for the hCOS20.6 sample. Circles are color coded by the corresponding D$_n4000$. The dashed grey curve in the top panel denotes the redshift evolution of the absolute magnitude limit corresponding to the magnitude limit of hCOS20.6 ($r_{lim} = 20.6$) for the star-forming (D$_n4000<1.5$) subsample. The limits $\log(M_\ast/M_\sun)=10.6$ and $z=0.4$ define an approximately mass complete sample. \label{f6}}
\end{centering}
\end{figure}

\begin{figure*}
\begin{centering}
\includegraphics[scale=0.475]{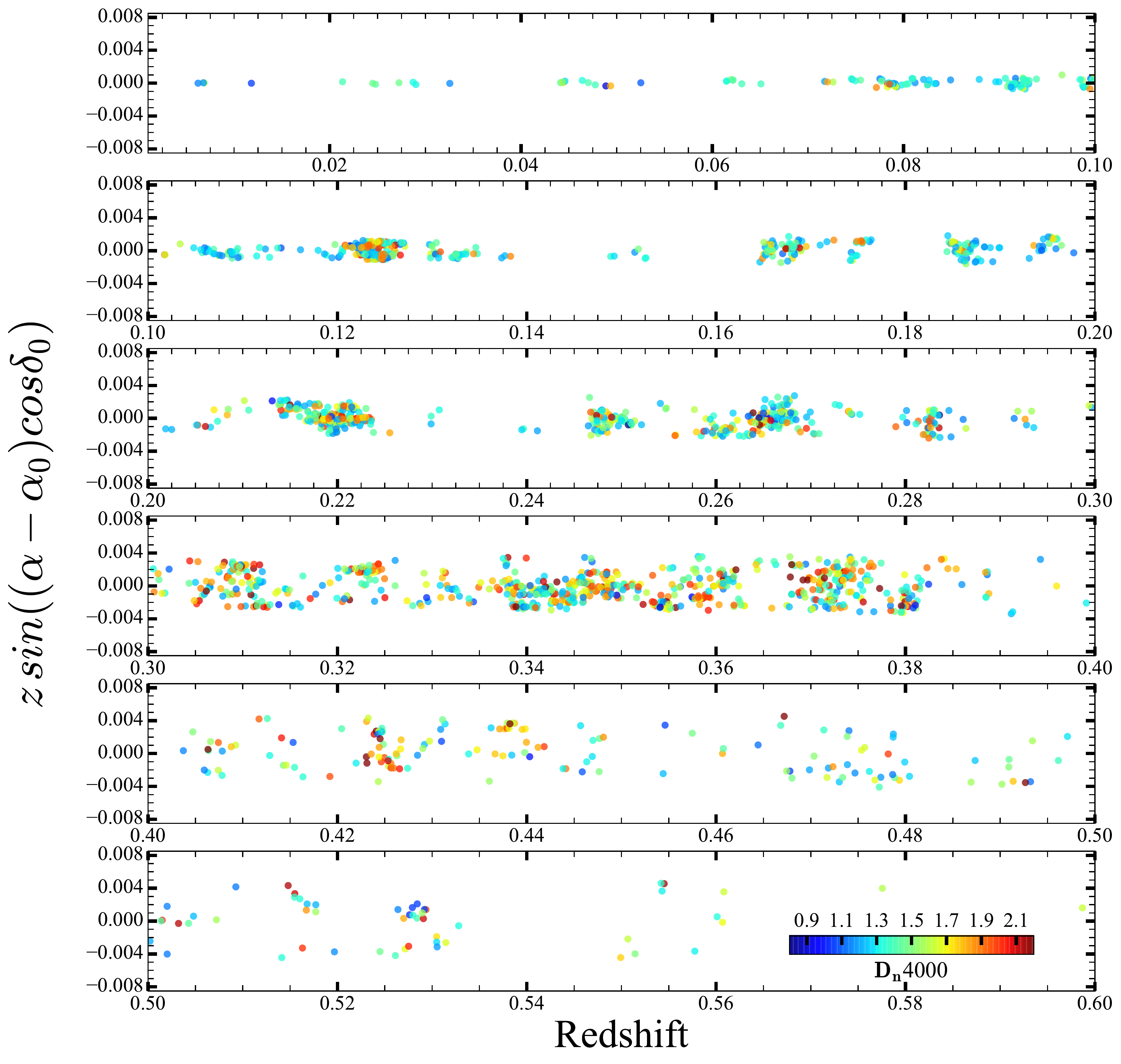}
\caption{ Cone digram for hCOS20.6 covering the central 0.89~deg$^2$ of the COSMOS field projected in R.A.$_\mathrm{2000}$. Galaxies are color-coded based on their D$_n4000$ index. \label{f7}}
\end{centering}
\end{figure*}

We compare the spatial and magnitude distributions of hCOSMOS galaxies and of the parent photometric UltraVISTA sample to construct a complete magnitude limited subsample, hCOC20.6,  within the central 0.89~deg$^2$ of the COSMOS field (see Table~\ref{table1}). We augment the hCOSMOS systems in hCOS20.6 with a small number (37) of SDSS/BOSS observations. In addition to the measurements for hCOSMOS sample, Table~\ref{table2} lists previously reported redshifts and our D$_n4000$, velocity dispersion, and stellar mass measurements for these  SDSS/BOSS targets.

The left-side panel of Figure~\ref{f5} shows the spectroscopic completeness of hCOS20.6 and zCOSMOS.  For $r<20.6$ the sample of 1968 hCOS20.6 galaxies is $>90\%$ complete. For comparison, we plot the zCOSMOS-bright completeness over the same magnitude range \citep{Lilly2007, Lilly2009}. The zCOSMOS-bright survey targets galaxies with $I_{AB}\leqslant22.5$ and is designed to provide a uniform completeness of $\sim60\%$ throughout this magnitude range for the full 1.7~deg$^2$ of the COSMOS field. In contrast, the hCOS20.6 survey provides $>90\%$ spectroscopic completeness in the central $\sim1$~deg$^2$ of the COSMOS field for objects with $r<20.6$~mag. The left-side panel of Figure~\ref{f5} reflects these differences in these complementary surveys.  

The hCOS20.6 sample is $\gtrsim90\%$ complete in every $6\arcmin\times6\arcmin$ spatial bin within the area outlined by the black solid line in the right-side panel of Figure~\ref{f5}. Uniform, dense spectroscopic surveys like hCOS20.6 are especially important for studies of the interplay between galaxy internal properties and their environments (e.g., the mapping of density fields around massive compact quiescent galaxies in hCOSMOS, \citealt{Damjanov2015b}) and for unbiased assessment of scaling relations among galaxy properties. 

The upper panel of Figure~\ref{f6} shows the absolute $r-$band magnitude as a function of redshift and  D$_n4000$ for hCOS20.6. We determine the $K-$corrected and reddening-corrected magnitude by fitting the stellar population synthesis models of \citet{Bruzual2003} to the observed SED using the {\sc Lephare} code just as we did to estimate galaxy stellar mass (Section~\ref{photo}). The grey dashed line in the upper panel of Figure~\ref{f6} traces the limiting absolute magnitude $M_{r,lim}$ as a function of redshift:

\begin{equation}
M_{r,lim}=m_{r,lim}-5\log\left(\frac{D_L(z)}{10\, \mathrm{pc}}\right)-\overline{K},
\end{equation}

where $m_{r,lim}=20.6$~mag is the limiting apparent magnitude for the complete sample, $D_L(z)$ is the luminosity distance, and $\overline{K}$ is the average $K$~correction for star-forming (D$_n4000<1.5$) galaxies as a function of redshift. A small number of star-forming galaxies (56, or 3\% of the complete sample) lie below the absolute magnitude limit. Galaxies scatter across the limit due to large photometric uncertainties.

In the lower panel of Figure~\ref{f6} we show the distribution of stellar masses  as a function of redshift and D$_n4000$. The stellar mass distribution displays the previously known trends: 1) at each redshift the most massive galaxies are have large D$_n4000$, i.e. they are dominated by old stellar populations, and 2) at a given stellar mass within the $9.5<\log(M_*\, [M_\sun])<11$ range the fraction of galaxies with small D$_n4000$ increases with redshift. These relations between stellar mass, galaxy quiescence, and redshift are apparent in other dense spectroscopic surveys covering a similar redshift interval (e.g., SHELS, \citealt{Geller2016, Geller2014} or PRIMUS, \citealt{Moustakas2013}).

Figure~\ref{f7} shows the distribution of hCOS20.6 galaxies in $\Delta z=0.1$ redshift bins projected along the R.A.$_\mathrm{2000}$ direction. The characteristic large-scale structure is clearly visible. Individual objects are color-coded by D$_n4000$. Figure~\ref{f7} shows that galaxies with small $D_n4000$ dominate lower density regions and galaxies with large $D_n4000$ are typically found in high density regions, again a known relation. 

\section{Properties of the magnitude-limited sample}\label{properties}

To demonstrate some  applications of hCOS20.6 and similar future surveys, we explore the distribution of D$_n$4000 indices and we examine the correlation between D$_n$4000 and galaxy rest-frame colors. We also compare the D$_n4000$ distribution of hCOS20.6 with the distribution  in the 4~deg$^2$ SHELS F2 field to the same limiting depth \citep{Geller2014}. Finally, we combine structural parameters based on HST imaging with galaxy stellar masses and D$_n4000$ indices to examine the size-stellar mass relation for star-forming and quiescent hCOS20.6 galaxies. 

\subsection{D$_n4000$: selecting the star-forming and quiescent galaxy populations}\label{dn4000_sfq}

\begin{figure*}
\begin{centering}
\includegraphics[scale=0.50]{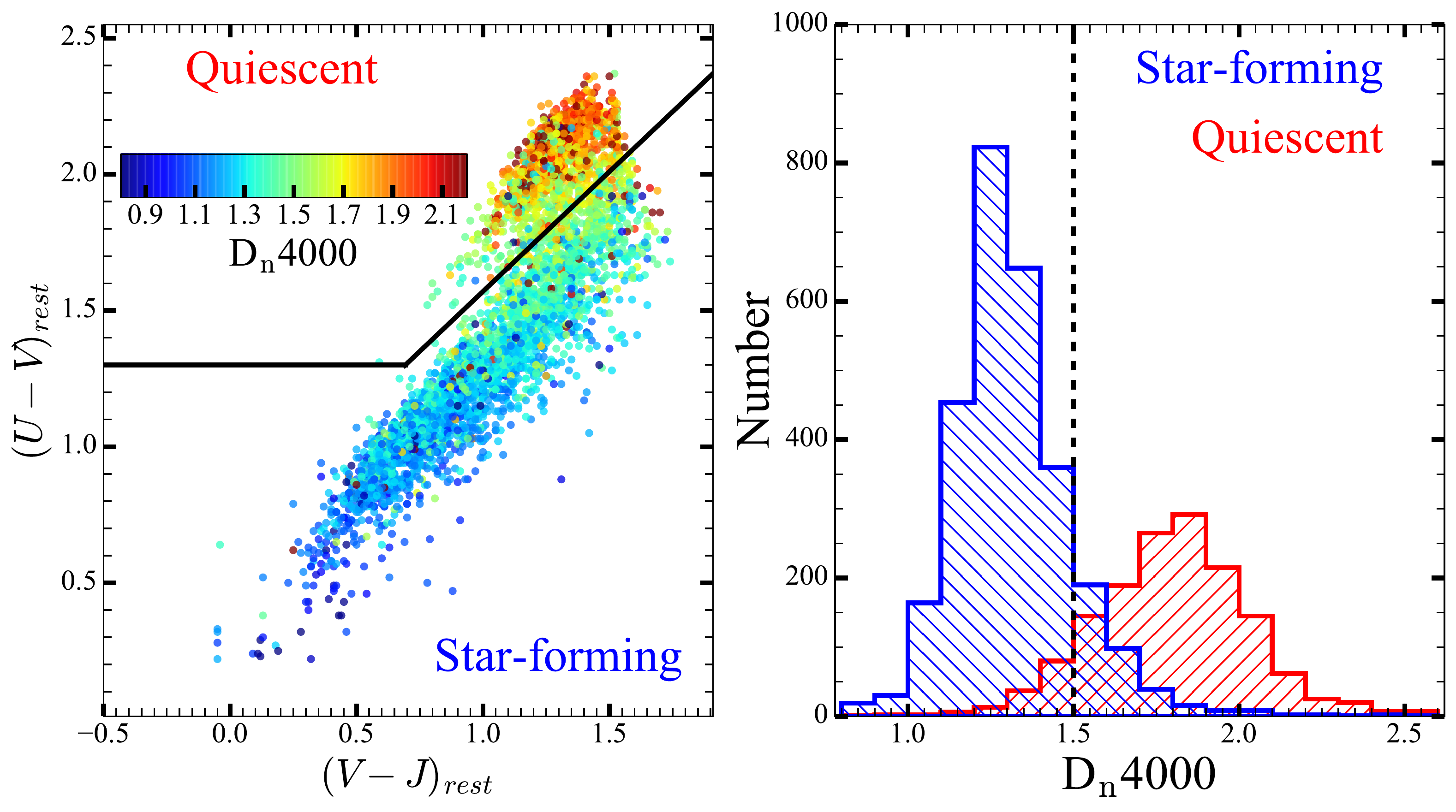}
\caption{Rest-frame UVJ color-color diagram for hCOS20.6 and SDSS galaxies with D$_n4000$ measurements (left). The color coding indicates the D$_n4000$ values. The separation between quiescent and star-forming galaxies (solid black line) is from \citet{Williams2009}. The distribution of D$_n4000$ for quiescent and star-forming galaxies selected based on their rest-frame $UVJ$ colors (right). \label{f8}}
\end{centering}
\end{figure*}

Large spectroscopic surveys demonstrate that the D$_n4000$ index distribution is strongly bimodal, with a clear division between quiescent and star-forming galaxies at D$_n4000\sim1.5$ \citep{Kauffmann2003, Vergani2008, Woods2010, Geller2014}. Throughout this work we define star-forming and quiescent galaxies as those with $D_n4000 < 1.5$ and $D_n4000 > 1.5$, respectively. Other methods for discriminating between star-forming and quiescent galaxies include galaxy morphology, observed colors, rest-frame colors, the shape of galaxy SEDs, and various combinations of photometric, spectroscopic, and/or morphological galaxy properties \citep[e.g.,][]{Moresco2013}. 

The $UVJ$ color-color diagram is one commonly used technique to separate star-forming and quiescent galaxies \citep{Williams2009}. The left-side panel of Figure~\ref{f8}  shows the $U-V$ versus $V-J$ rest-frame colors of hCOS20.6 galaxies. The solid black line shows the  separation between star-forming and quiescent galaxy populations from \citet{Williams2009}. Galaxies in the color-color diagram are color-coded by their D$_n4000$ index, demonstrating that the large majority of D$_n4000>1.5$ systems (yellow/red points) occupy the quiescent population region  defined by galaxy rest-frame $(U-V)$ and $(V-J)$ colors. In contrast, galaxies with D$_n4000<1.5$ (blue and green points) have rest-frame colors of star-forming systems. 

No technique can separate star-forming and quiescent galaxies perfectly. Some mixing of two galaxy population is expected for all classifiers \citep[e.g.,][]{Woods2010, Moresco2013}. We compare the  identification of quiescent and star-forming samples on the basis of their rest-frame colors with the identification based on D$_n4000$ . The blue and red histograms in the right-side panel of Figure~\ref{f8}) show the distributions of D$_n$4000 for the quiescent and star-forming populations identified on the basis of rest-frame color.  In the passive rest-frame color selected sample (red histogram) less than $10\%$ of galaxies have D$_n4000<1.5$. For the color-selected star-forming population $\sim13\%$ are quiescent based on their spectroscopic properties (D$_n4000>1.5$). We note that the contamination levels reported here are lower than any of the classifiers tested in \citet{Moresco2013}. We thus confirm and extend the results based on the quiescent hCOSMOS sample selection \citep{Damjanov2015b}: for galaxies {\it with measured spectroscopic redshifts} the D$_n4000$ spectroscopic indicator and the rest-frame colors provide very similar samples of quiescent and star-forming systems.

\subsection{Large-Scale Structure and D$_n4000$: comparison with the F2 sample}\label{hCOS-F2}

\begin{figure}
\begin{centering}
\includegraphics[scale=0.35]{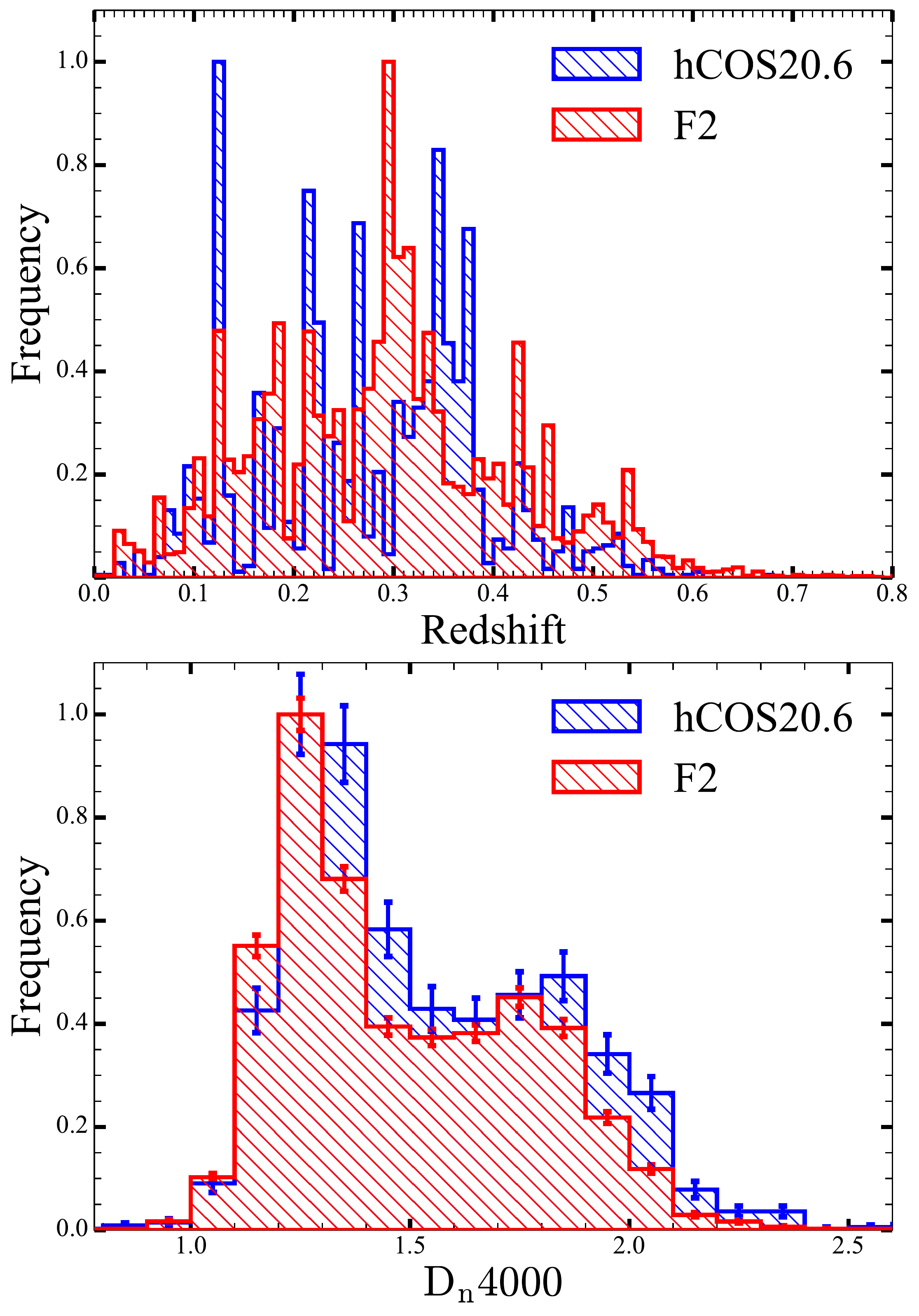}
\caption{ Distribution of redshift (upper panel) and D$_n4000$  (lower panel)  for two complete magnitude limited spectroscopic samples: hCOS20.6 (blue lines)  and F2 \citep[][ red lines, 4 square degree area limited to r = 20.6]{Geller2014}. Error bars in the lower panel represent Poisson errors. \label{f9}}
\end{centering}
\end{figure}

The F2 field is one of two 4~deg$^2$ fields of  SHELS (Smithsonian Hectospec Lensing Survey), a dense complete redshift survey to a limiting $R-$band magnitude $R=20.6$ \citep{Geller2014, Geller2016}. The $95\%$ complete survey of the F2 field, centered at R.A.$_\mathrm{2000}=09^\mathrm{h}19^\mathrm{m}32^\mathrm{s}$ and Dec.$_\mathrm{2000}=+30\arcdeg 00\arcmin 00\arcsec$, includes $\sim13,300$ objects with spectroscopic redshift, stellar mass, and D$_n4000$ index measurements. For the comparison with hCOS20.6, we construct a magnitude limited ($r<20.6$) subsample of F2 galaxies using the $R$~to~$r$ conversion provided in \citet[Eq.~1]{Geller2016}.

The upper panel of Figure~\ref{f9} shows the redshift distribution for galaxies in the hCOS20.6 (blue) and F2 (red) samples; the distributions clearly differ. The F2 sample describes the structure in the field in the $0.1\lesssim z\lesssim0.6$ redshift range, with the median redshift $\widetilde{z}=0.3$. The hCOS20.6 sample follows the structure of the COSMOS field in the redshift interval $0.1\lesssim z\lesssim0.4$, with the median value of $\widetilde{z}=0.27$ (as shown in the the cone diagram, Figure~\ref{f8}). The richest structures in the F2 field are concentrated around $z\sim0.3$, followed by prominent peaks at $z\sim0.12$, $z\sim0.18$, $z\sim0.21$, and $z\sim0.42$. In the hCOS20.6 sample, a large fraction of the galaxies are concentrated at $z\sim0.12$, with additional peaks in the redshift distribution at $z\sim0.22$, $z\sim0.26$, and $z\sim0.35$. Differences in the redshift distributions reflect  cosmic variance.

The lower panel of Figure~\ref{f9} shows the D$_n4000$ distribution for galaxies in hCOS20.6 (blue) and F2 (red). In striking contrast to the redshift distributions, the F2 and hCOS20.6 samples show very similar D$_n4000$ distributions. The two-sample Kolmogorov-Smirnov (K-S) test gives a $p-$value of 0.01 for the redshift distributions in the two fields; the hypothesis that the two distributions originate from the same parent distribution is rejected. The $p-$value of the two-sample K-S test performed on the two D$_n4000$ distributions is much higher ($p=0.11$); thus we cannot reject the hypothesis that the distributions of D$_n4000$ in F2 and hCOS20.6 have the same parent distribution. We obtain very similar results with the Anderson-Darling two-sample test \citep{Scholz1987} that is more sensitive to the tails of distributions than the K-S test. 

Comparison of the D$_n$4000 distributions for the hCOS20.6 and F2 samples reveals that, averaged over a broad redshift range that includes both dense structures and low density regions, the fractions of quiescent (D$_n4000>1.5$) and star-forming (D$_n4000<1.5$) galaxies are consistent between the two fields \citep[see also][] {Geller2016}. Although hCOS20.6 survey is small and thus subject to substantial cosmic variance, the distribution of D$_n4000$ indices is remarkably similar to the SHELS F2 survey that covers $\gtrsim5\times$ larger area to a similar depth. This comparison demonstrates that in spite of cosmic variance, the distribution of D$_n4000$ index is stable when averaged over volumes at least as large as hCOS20.6.

\subsection{D$_n4000$ index: galaxy size - stellar mass relation}\label{size-mass}

\begin{figure}
\begin{centering}
\includegraphics[scale=0.35]{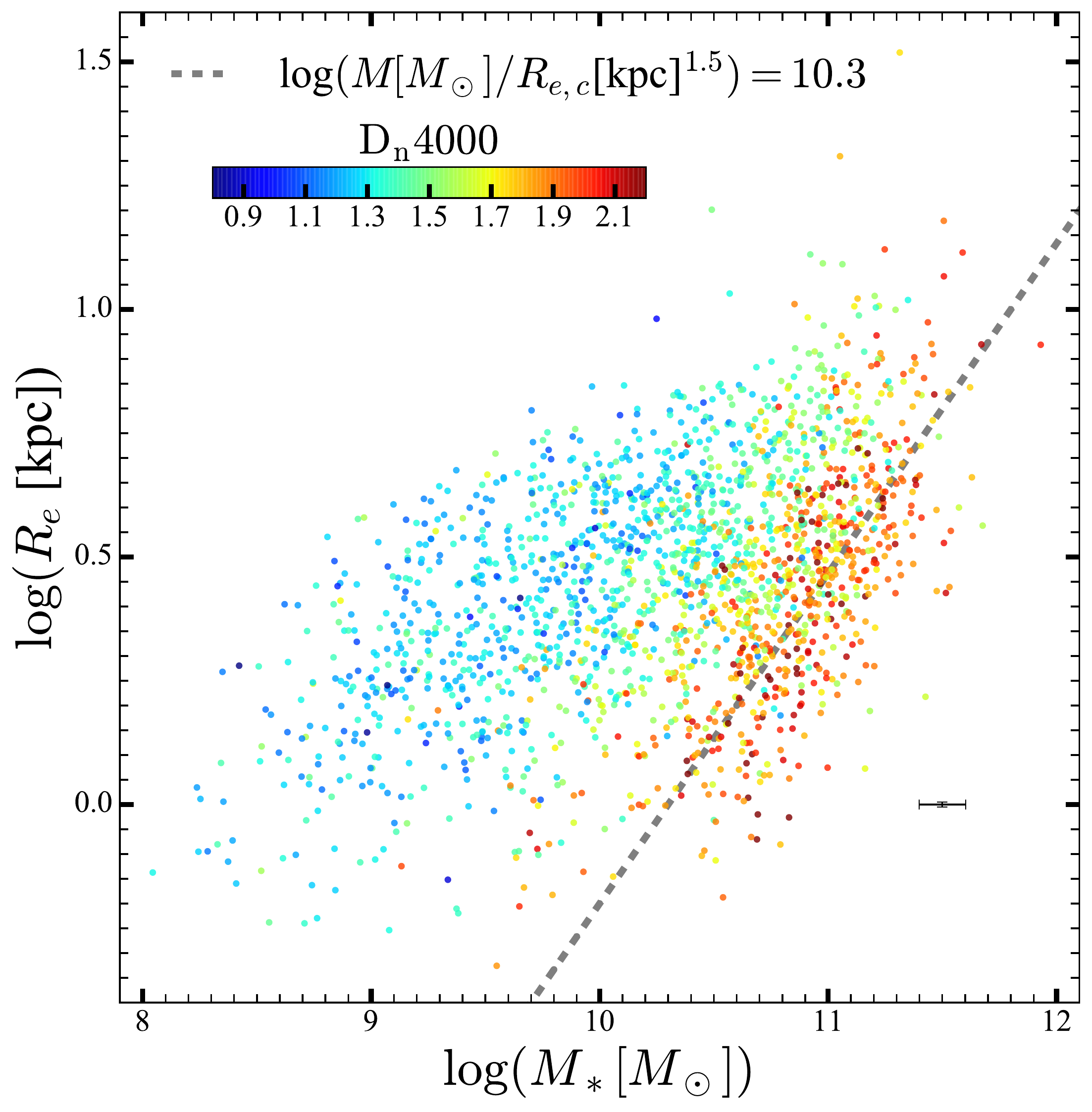}
\caption{Circularized effective radius as a function of stellar mass for hCOS20.6. The circles are color coded by D$_n4000$. The dashed line shows the compactness cutoff from \citet{Barro2013}. Singe error bars on the bottom right represent typical errors.\label{f10}}
\end{centering}
\end{figure}

Both star-forming and quiescent galaxy populations exhibit a correlation between  stellar mass and size. In recent years many studies have explored on the size - stellar mass relation and its redshift evolution \citep[e.g.,][]{Shen2003, Trujillo2004, Maltby2010, vanDokkum2010, Damjanov2011, Newman2012, HuertasCompany2013, vanderWel2014, Delaye2014, PaulinoAfonso2017}. The accuracy of the relation depends on the range of stellar masses, errors in the sizes and stellar masses, the resolution of the images used to estimate galaxy sizes, surface brightness selection effects, the imaging rest-frame wavelength, and the range of environments \citep[e.g,][]{Lange2015, vanderWel2014, Sweet2017}. Thus the size - stellar mass relation can vary significantly between different galaxy samples. However, in comparison with their quiescent counterparts, star-forming galaxies at all redshifts exhibit a shallower trend in size with stellar mass (see e.g. \citet{Shen2003} for $z\sim0$ relations and \citet{vanderWel2014} for $0.5<z\lesssim3$ relations).

Figure~\ref{f10} shows that the star-forming and quiescent populations of hCOS20.6 follow different size - stellar mass relations, $\log(R_e\, [\mathrm{kpc}])\propto \log(M_*\, [M_\sun])^\alpha$. A detailed analysis of the size-stellar mass relation is beyond the scope of this work. However, we note that the size of quiescent hCOS20.6 systems displays a trend with stellar mass ($\alpha\sim0.45$) that is steeper than the one for star-forming galaxies in our sample ($\alpha\sim0.22$), in qualitative agreement with other recent studies \citep[e.g,][]{Haines2017, Lange2016, vanderWel2014}. Furthermore, the slopes of the size-mass relations for quiescent and star-forming hCOS20.6 galaxies agree within the uncertainties ($<2\sigma$) with the size-mass relations that we derive for the two galaxy populations in the much larger SHELS F2 sample based on Subaru Hyper Suprime-Cam images (Damjanov et al. in prep).

The size - stellar mass diagram of Figure~\ref{f10} can  be used to select samples of compact galaxies as the extreme outliers from the size - stellar mass relation defined by the parent star-forming or quiescent population \citep[e.g.,][]{Trujillo2007, vanDokkum2008, Poggianti2013, Barro2013, vanderWel2014}. As in our previous analysis of the COSMOS field \citep{Damjanov2015a, Damjanov2015b, Zahid2015, Zahid2016a},  we employ the threshold for galaxy compactness defined by \citet[][grey dashed line in Figure~\ref{f10}]{Barro2013}. Based on this selection, 249 hCOS20.6 galaxies (or $\sim13\%$ of the parent sample) are compact.  In addition, the color-coding of hCOS20.6 systems in Figure~\ref{f10} shows that the compact systems in our $0.1<z\lesssim0.6$ sample tend to be quiescent (D$_n4000\geqslant1.5$). 

\begin{figure}
\begin{centering}
\includegraphics[scale=0.25]{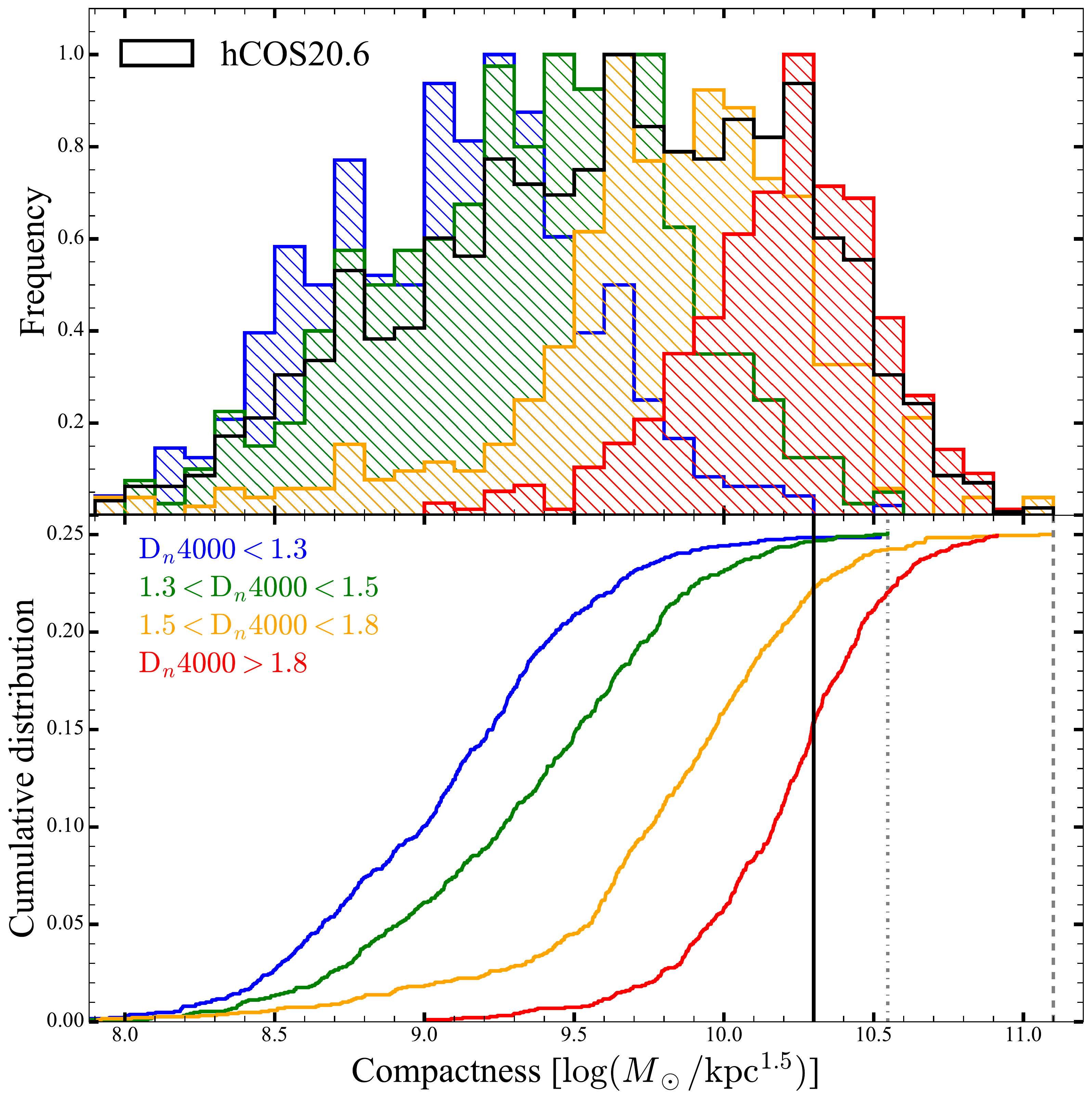}

\caption{Upper panel: Distribution of compactness for hCOS20.6 (black) and for subsamples spanning different D$_n4000$ ranges (blue, green, orange, red). Lower panel: Fractional cumulative distribution of compactness for the four subsamples. The solid black line corresponds to the compactness cutoff from Figure~\ref{f10}. The dash-dotted and dashed grey lines show the maximum compactness for the star-forming (D$_n4000<1.5$) and quiescent (D$_n4000>1.5$) hCOS20.6 subsamples, respectively. \label{f11}}
\end{centering}
\end{figure}

We explore the relation between galaxy compactness and age (quiescence) in more detail in Figure~\ref{f11}. Here we select four galaxy subsamples in equally populated bins of D$_n4000$. Normalized distributions of galaxy compactness (defined as in Figure~\ref{f10}) show that hCOS20.6 systems with the highest D$_n4000$  are also the most compact (red histogram in the upper panel of Figure~\ref{f11}). Furthermore, there is a very little overlap between the compactness distributions for galaxies with the oldest (D$_n4000>1.8$) and the youngest (D$_n4000<1.3$) stellar population. Galaxies with the lowest D$_n4000$ indices are the least compact ones in the parent sample  (blue histogram in the upper panel of Figure~\ref{f11}). 

Cumulative distributions in the lower panel of Figure~\ref{f11} illustrate even more clearly the strong positive trend in compactness with the dominant stellar population age for the galaxy. Maximum compactness levels of the four hCOS20.6 subsamples (grey dashed and dash-dotted lines in the lower panel of Figure~\ref{f11}) demonstrate that quiescent galaxies can be up to a factor of four times more compact than the star-forming objects. Note that we do not perform this comparison at fixed stellar mass. Figure~\ref{f10} shows that in our magnitude limited sample star-forming galaxies dominate the low stellar mass region and quiescent systems occupy the high-mass end of the stellar mass distribution. Thus the trend in compactness with galaxy age (Figure~\ref{f11}) is driven mainly by the large stellar mass of galaxies with high D$_n4000$.  

\subsection{Velocity dispersion: relation between stellar mass and velocity dispersion} 

\begin{figure}
\begin{centering}
\includegraphics[scale=0.85]{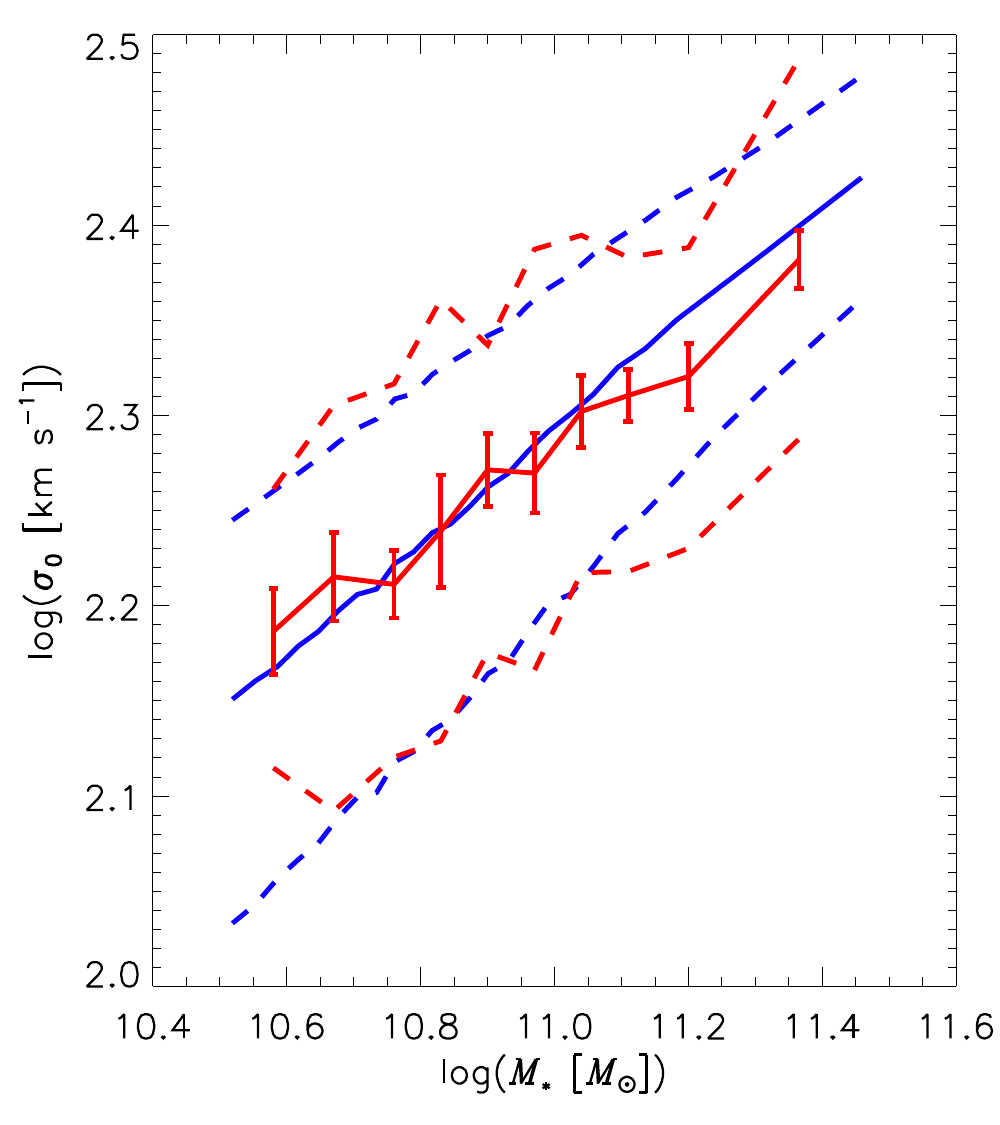}
\caption{Relation between velocity dispersion and stellar mass. Solid red and blue curves show the relation for hCOS20.6 and SDSS, respectively. The relation is determined by taking the median velocity dispersion in equally populated bins of stellar mass. Error bars are from bootstrapping the data. The dashed red and blue curves are the intrinsic scatter of velocity dispersion as a function of stellar mass for hCOS20.6 and SDSS, respectively. We derive the intrinsic scatter empirically by measuring the limits of the central 68\% of the velocity dispersion distribution and subtracting the median observational uncertainty in quadrature for each stellar mass bin respectively. Note the lack of significant evolution in both the relation and intrinsic scatter. \label{f12}} 
\end{centering}
\end{figure}

The central stellar velocity dispersion of quiescent galaxies is correlated with stellar mass---the $M_\ast - \sigma$ relation. We derive the $M_\ast - \sigma$ relation for hCOS20.6 sample using an analysis approach similar to \citet{Zahid2016b}. We select quiescent galaxies with D$_n4000 > 1.5$, $0.2 < z< 0.5$ and $\log(M_\ast/M_\sun) > 10.5$; the mass selection mitigates limitations due to instrumental resolution. These selection criteria yield a sample of 597 quiescent galaxies. We have a velocity dispersion measurement for 95\% of galaxies in this selected sample (565/597). We aperture correct the velocity dispersion to a fiducial physical aperture radius of 3 kpc (see \citealt{Zahid2016b} for details). The aperture correction is small and does not significantly affect the results. We compare the hCOS20.6 data to a consistently analyzed local sample from SDSS. Details of the SDSS sample selection are in \citet{Zahid2016b}.

Figure~\ref{f12} shows the $M_\ast  - \sigma$ relation for hCOS20.6 and SDSS galaxies. We fit the two relations using \emph{linfit.pro} in IDL. The best fit is 

\begin{equation}
\mathrm{log}(\sigma) \left( M_{11} \right) = (2.285 \pm 0.006) + (0.25 \pm 0.03) M_{11}
\label{eq:hcos_fit}
\end{equation}
and
\begin{equation}
\mathrm{log}(\sigma) \left( M_{11} \right) = (2.2926 \pm 0.0002) + (0.298 \pm 0.001) M_{11}
\label{eq:sdss_fit}
\end{equation}
for hCOS20.6 and SDSS, respectively. Here $M_{11} = \log(M_\ast [M_\sun]) - 11$. The two relations are consistent. From Figure~\ref{f12} we conclude that the $M_\ast  - \sigma$ relation of massive quiescent galaxies does not evolve signficantly for galaxies at $z<0.5$ and that the intrinsic scatter does not depend strongly on redshift. 

 \citet{Zahid2016b} examine the $M_\ast - \sigma$ relation derived from a sample of $\sim 4500$ galaxies in the SHELS-F2 at $z\lesssim0.7$. They sort their sample in bins of redshift and find no significant evolution in the $M_\ast - \sigma$ relation. They measure $\sigma \propto M_\ast^{\sim0.3}$. Results in Figure~\ref{f12} are consistent with the detailed analysis of the $M_\ast - \sigma$ relation presented in \citet{Zahid2016b}. The kinematic scaling relations measured from hCOSMOS are robust and demonstrate the power of these data to explore the spectral kinematic properties of quiescent galaxies at intermediate redshifts.

\section{Testing photometric proxies for the central velocity dispersion}\label{proxy}

\begin{figure}
\begin{centering}
\includegraphics[scale=0.35]{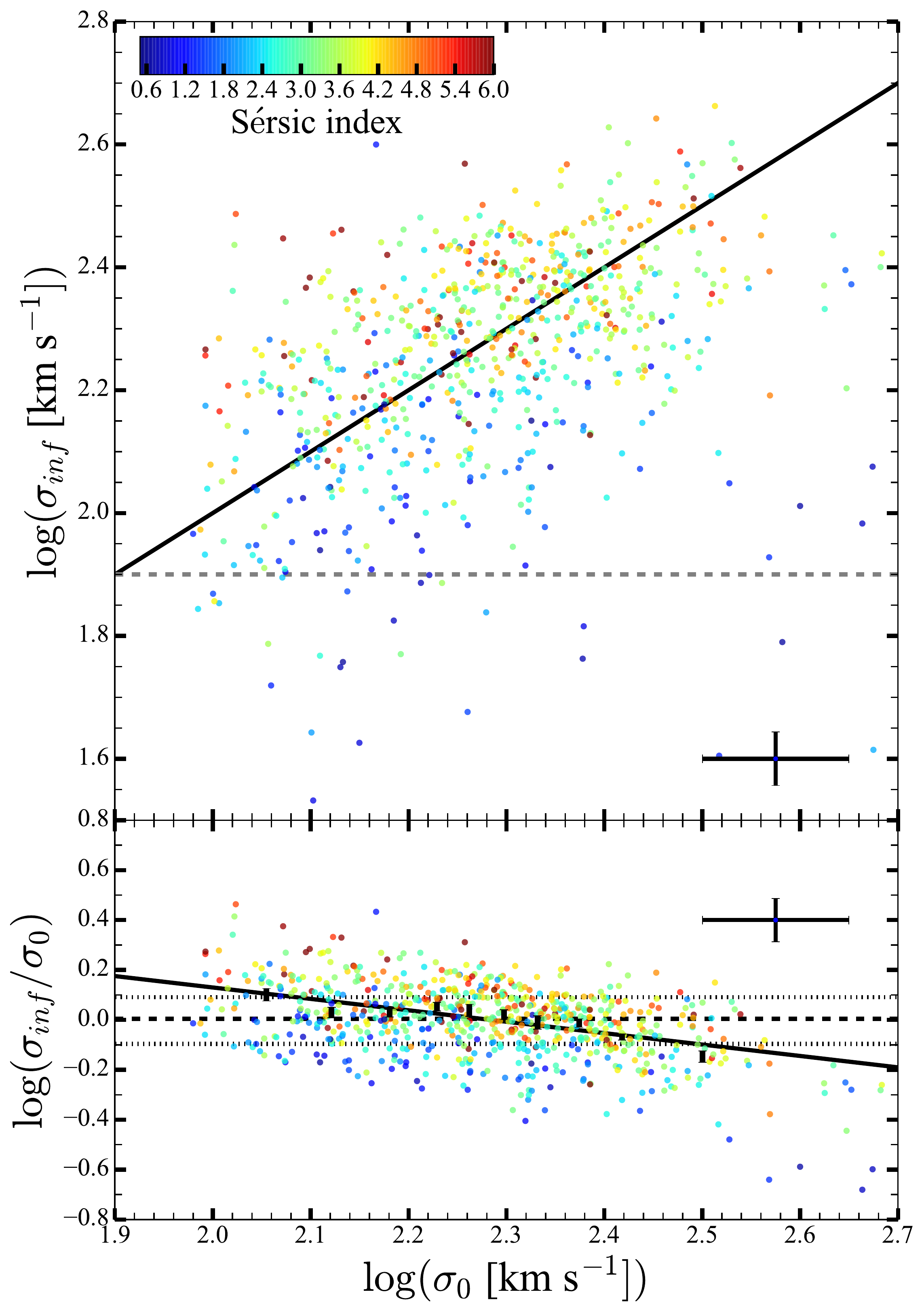}
\caption{Upper panel: inferred central velocity dispersion vs. measured central velocity dispersion. The points are color-coded by S\'ersic index.The black solid line is the 1:1 relation; the grey dashed line correspond to the $\sigma_{inf}>80$~km~s$^{-1}$ threshold for the inferred velocity dispersion. Lower panel: the ratio between inferred and measured velocity dispersions vs. measured velocity dispersion. The dashed and dotted  lines show the median and the interquartile range for the sample. Black points with error bars represent the median value and its bootstrapped error for $\sigma_{inf}/\sigma_{0}$ in equally populated bins of measured velocity dispersion $\sigma_0$. The black solid line shows the best fit to the trend. Single bars in both panels represent typical errors. \label{f13}}
\end{centering}
\end{figure}

\begin{figure*}
\begin{centering}
\includegraphics[scale=0.35]{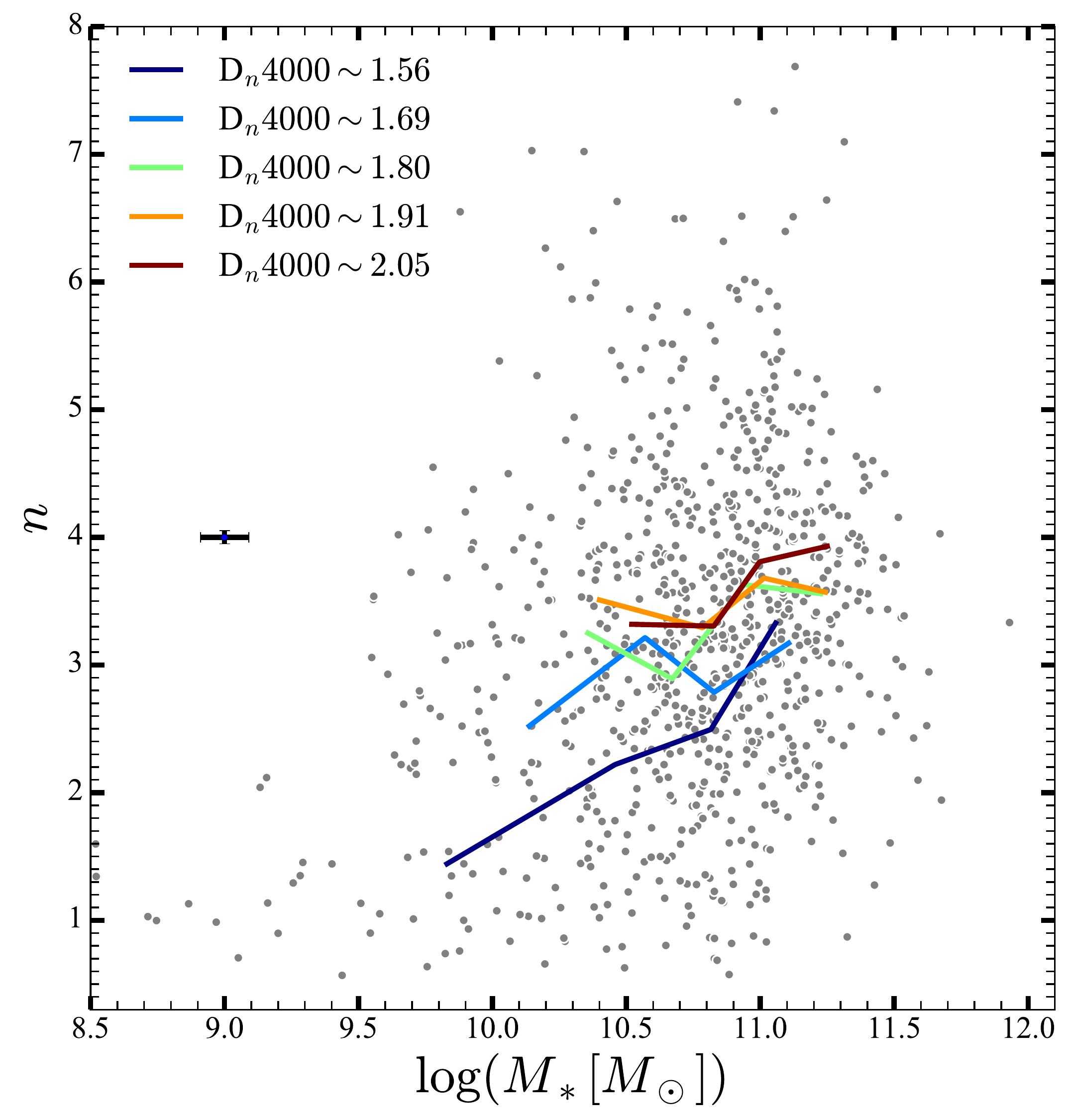}
\includegraphics[scale=0.35]{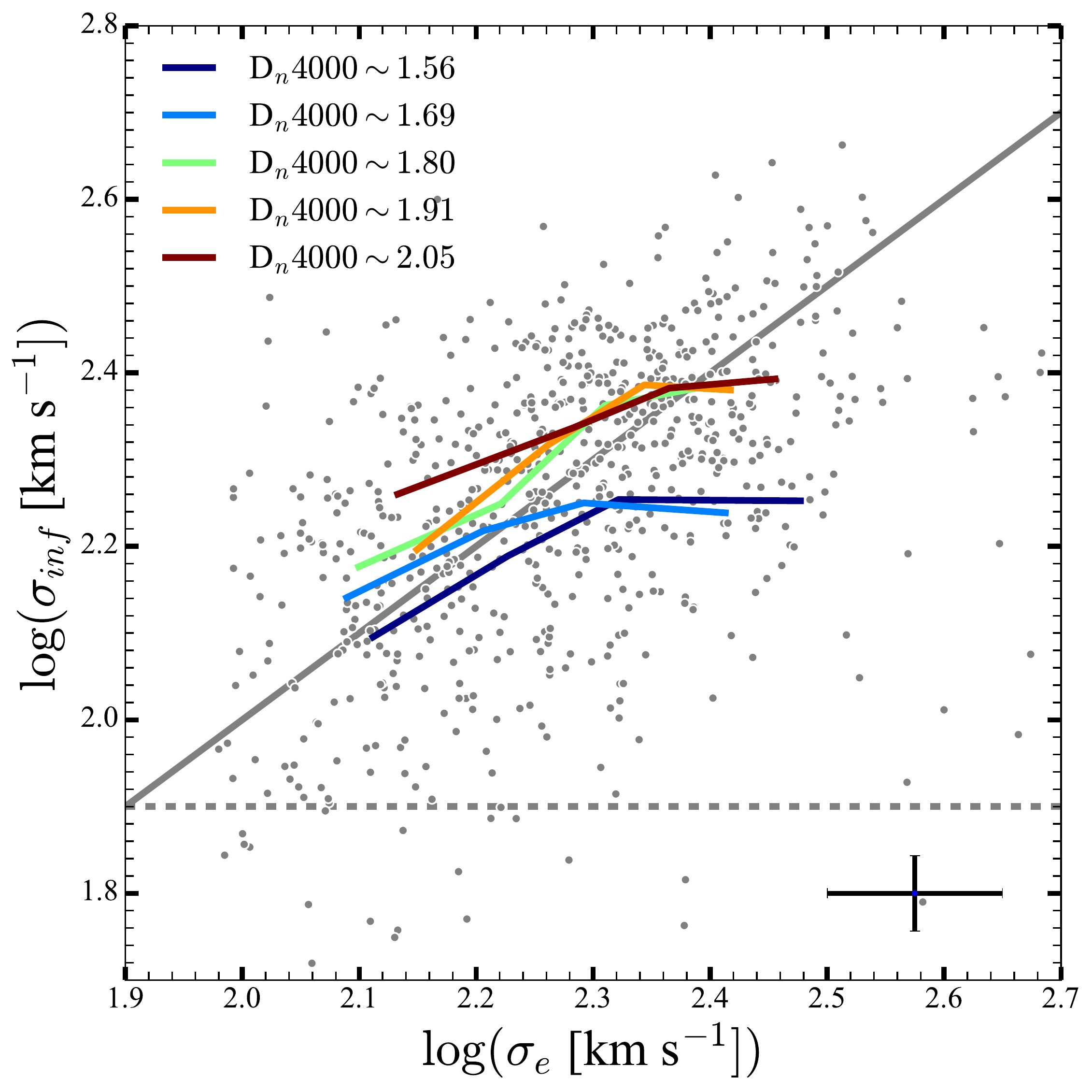}

\caption{Left: Median S\'ersic index as a function of stellar mass and D$_n4000$ for the quiescent (D$_n4000>1.5$) galaxies in the hCOS20.6 sample. Colored solid lines follow the median S\'ersic index in bins of stellar mass for galaxies with a range of D$_n4000$ measurements. Right: median inferred velocity dispersion as a function of measured velocity dispersion and D$_n4000$ of quiescent hCOS20.6  galaxies. Colored solid lines follow the median inferred velocity dispersion in bins of measured velocity dispersion for galaxies with a range of D$_n4000$. Grey dashed line shows the threshold for inferred velocity dispersion from Figure~\ref{f13}. Grey circles in both panels correspond to individual galaxies. Single error bars in both panels represent typical errors.\label{f14}}
\end{centering}
\end{figure*}

\begin{figure}
\begin{centering}
\includegraphics[scale=0.35]{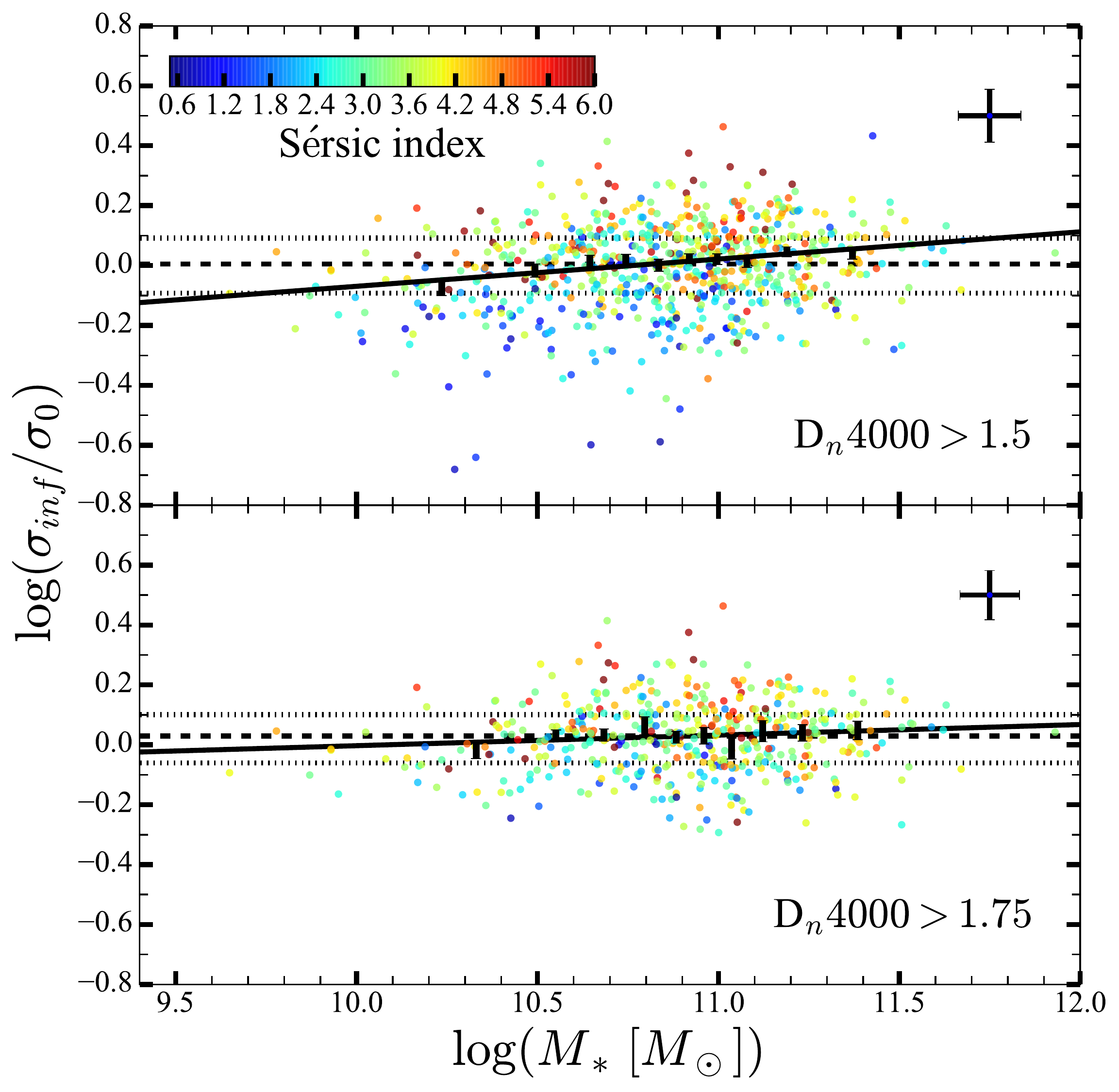}
\caption{The ratio between predicted and measured velocity dispersions vs. stellar mass for quiescent hCOS20.6 galaxies (upper panel: all D$_n4000>1.5$ systems with measured velocity dispersions; lower panel: the D$_n4000>1.75$ subsample). In both panels points are color-coded by S\'ersic index and dashed and dotted  lines show the median and the interquartile ranges for the two samples. Black points with error bars represent the median value and its bootstrapped error for $\sigma_{inf}/\sigma_{0}$ in equally populated bins of stellar mass. Black solid lines in both panels corresponds to the best fit to the trend that median 
$\sigma_{inf}/\sigma_{0}$ values show with stellar mass. Single error bars in both panels represent typical errors. \label{f15}}
\end{centering}
\end{figure}

Velocity dispersion is correlated with stellar mass, size and Sersic index \citep[e.g,][]{Taylor2010, Bezanson2011, Zahid2017}. \citet{Bezanson2011} calibrate a photometric proxy for stellar velocity dispersion using the SDSS DR7 galaxy sample \citep[see also][]{Bezanson2012}. They apply this proxy to construct the velocity dispersion function and to examine the evolution of galaxy dynamical properties at $0<z<1.5$. We use the quiescent hCOS20.6 galaxies with measured velocity dispersions to test this proxy at $0.05<z<0.66$.

We compare the measured velocity dispersion $\sigma_0$ with the value inferred from photometry \citep{Bezanson2011}:

\begin{equation}
\sigma_{inf} =\sqrt{\frac{GM_\ast}{0.577K_v\left(n\right)R_{e,c}}} \label{eq:def}.
\end{equation}

Here the 0.577 factor is the average stellar-to-total mass ratio $\langle M_\ast/M_{dyn}=0.577\rangle$ derived by \citet{Bezanson2011}. The coefficient $K_v$ in Equation~\ref{eq:def} is related to the S\'ersic index $n$ \citep{Bertin2002}:

\begin{equation}\label{eq:kv}
K_v(n)=\frac{73.32}{10.465+{\left(n-0.94\right)}^2} + 0.954.
\end{equation}

\noindent This relation is derived assuming virial equilibrium.

To estimate the inferred velocity dispersion using Equations~\ref{eq:def}~and~\ref{eq:kv} we combine our stellar mass estimates for the hCOSMOS sample with \citet{Sargent2007} GIM2D-based galaxy size and S\'ersic index measurements (see Section~\ref{photo} for more details on these measurements).

Velocity dispersion measured from galaxy spectra depends on the aperture size. To correct our hCOSMOS velocity dispersion measurements ($\sigma_m$) to an aperture radius of $r_{e,c}/8$ \citep{Jorgensen1995, Bertin2002}, we use the form 

\begin{equation}
\frac{\sigma\left(\frac{r_{e,c}}{8}\right)}{\sigma_{m}}\equiv\frac{\sigma_0}{\sigma_{m}}=\left(\frac{\frac{r_{e,c}}{8}}{0\farcs75}\right)^{-0.033}\label{eq:apcor}, 
\end{equation}
 
where 0\farcs75 is the size of the fiber aperture for Hectospec. This correction results from the comparison between velocity dispersion measurements based on SDSS and Hectospec spectra for SHELS galaxies \citep{Zahid2016b}. The Hectospec-based correction is consistent with aperture corrections based on observations with other fiber-fed spectrographs \citep{Jorgensen1995}, long-slit spectrographs \citep{Mehlert2003}, and integral field units \citep{Cappellari2006}.

The upper panel of Figure~\ref{f13} compares the inferred ($\sigma_{inf}$) with the measured ($\sigma_{0}$) velocity dispersions color-coded by S\'ersic index. The root-mean-square scatter around the one-to-one relation (excluding outliers) is 0.14~dex ($\sim38\%$). \citet{Belli2014} report a similar scatter (0.13~dex or $\sim35\%$) based on a small sample of massive ($M_\ast>2.5\times10^{10}\, M_\sun$) quiescent galaxies at $0.9<z<1.6$. This scatter is larger than the typical 10\% uncertainty of our velocity dispersion measurements; thus the difference is dominated by systematic rather than statistical uncertainty. The upper panel of Figure~\ref{f13} shows that the difference depends on galaxy shape. Quiescent hCOSMOS galaxies with S\'ersic indices $n\lesssim2.5$ show the largest discrepancies between measured and inferred velocity dispersion.  

The lower panel of Figure~\ref{f13} shows the ratio between the inferred and measured velocity dispersion as a function of the measured central velocity dispersion. The median value of the ratio between two quantities is $\sigma_{inf}/\sigma_{0}=1.01$. The difference between the measured and inferred velocity dispersion is correlated with the velocity dispersion measurement (with the Pearson's correlation coefficient of $r_P=-0.51$). The best linear fit to the median residuals for equally populated bins of measured velocity dispersion (solid black line and black squares, respectively, in the lower panel of Figure~\ref{f13}) is

\begin{equation}
\frac{\sigma_{inf}}{\sigma_{0}}=(1.0\pm 0.2)\times\left(\frac{\sigma_{0}}{1\, \mathrm{km\, s}^{-1}}\right)^{(-0.46\pm 0.08)}.
\end{equation}

Figure~\ref{f13} shows that the residuals depend on galaxy shape (described by the S\'ersic index $n$). The slopes of the $\sigma_{inf}/\sigma_{0}$ vs. $\sigma_0$ relation for a subsample of quiescent galaxies with disk-like shapes ($n<2.5$) agrees (within uncertainties) with the slope of the relation for $n>2.5$ galaxies (spheroids). However, the offset between two relations is $\sim0.15$~dex for the measured velocity dispersion range $2\leqslant\log(\sigma_0)\leqslant2.5$. The dependence of the residuals on galaxy shape suggests that $K_v(n)$ (Eq.~\ref{eq:kv}) does not fully account for the dependence of velocity dispersion on S\'ersic index \citep[see e..g,][]{Zahid2017}.
 
To investigate the relation between S\'ersic index and inferred inferred velocity dispersion, we explore how this relation depends on the D$_n4000$ index. The left-side panel of Figure~\ref{f14} shows galaxy S\'ersic index as a function of its stellar mass. Colored lines connect median S\'ersic indices in equally populated stellar mass bins for different intervals of D$_n4000$ values. At high D$_n4000$ values (D$_n4000\gtrsim1.85$) the median S\'ersic index is not a strong function of stellar mass ($3<\widetilde{n}<4$). In contrast, at D$_n4000\lesssim1.6$ the median S\'ersic index varies greatly with stellar mass. An approximately mass-complete sample of $\log(M_*/M_\sun)>10.6$ quiescent hCOS20.6 galaxies at $z<0.4$ shows a similar trend. Furthermore, \citet{Zahid2017} find a qualitatively similar trend using a large sample of quiescent SDSS and SHELS galaxies at $z<0.7$ (see their Figure~6). As with the quiescent hCOS20.6 galaxies, they find that the trend in $\widetilde{n}$ with stellar mass flattens with increasing D$_n4000$.    

The relation between S\'ersic index, stellar mass, and D$_n4000$ implies a relation between the accuracy of the inferred velocity dispersion and D$_n4000$ index (the right-side panel of Figure~\ref{f14}). For galaxies with D$_n4000\lesssim1.65$ the inferred velocity dispersion is typically underestimated (e.g, for galaxies with $\sigma_0$ $\gtrsim200$~km~s$^{-1}$, $\widetilde\sigma_{inf}\sim0.6\times\widetilde\sigma_0$). For galaxies with higher D$_n4000$ index values (D$_n4000>1.75$) the median inferred velocity dispersion is overestimated for $\sigma_0<230$~km~s$^{-1}$ and $\sigma_{inf}$ accurately reproduces $\sigma_0$ (i.e, is within $10\%$ of the measured value) only for a narrow range of $\sigma_0$. 

Figure~\ref{f15} shows $\sigma_{inf}/\sigma_0$ as a function of stellar mass. The ratio is weakly correlated with stellar mass ($r_P^2\sim0.3$). The best fit relation is

\begin{equation}\label{sigma-mass}
\frac{\sigma_{inf}}{\sigma_{0}}=(-1.0 \pm 0.2)\times\left(\frac{M_\ast}{1\, M_\sun}\right)^{(0.09 \pm 0.02)}. 
\end{equation}

This relation (solid black line in the upper panel of Figure~\ref{f15}) is the best fit to the median values of $\sigma_{inf}/\sigma_0$ in equally populated stellar mass bins (black squares in the upper panel of Figure~\ref{f15}).  hCOS20.6 quiescent systems with $n<2.5$ (blue circles in the upper panel of Figure~\ref{f15}) show the largest offset from the median $\sigma_{inf}/\sigma_0$  (dashed line in the upper panel of Figure~\ref{f15}).

To test the effect of D$_n4000$  on the relation between the $\sigma_{inf}/\sigma_0$ ratio and stellar mass, we select a subsample of galaxies with D$_n4000>1.75$.  Figure~\ref{f14}  shows that galaxies with D$_n4000>1.75$ have little variation in median S\'ersic index  with increasing stellar mass. Consequently, the systems with large D$_n4000$ display median ratios between inferred and measured velocity dispersions (points with vertical error bars in the lower panel of Figure~\ref{f15}) that  consistently exceed one for $M_\ast\gtrsim2\times10^{10}\, M_\sun$. However, the trend in $\sigma_{inf}/\sigma_0$ with stellar mass for D$_n4000>1.75$ systems is consistent (within uncertainties) with the slope of the relation for all quiescent hCOS20.6 galaxies (Eq~\ref{sigma-mass}). 

The velocity dispersion inferred from galaxy photometric properties differs systematically from the directly measured value.  The discrepancy exceeds the statistical uncertainties in the measurements and the deviations are correlated with  galaxy properties. Systematic differences depend on dynamical parameters (the measured velocity dispersion) and stellar population properties (D$_n4000$ index, stellar mass).  We caution that systematic uncertainties in the photometric proxy for velocity dispersion may introduce spurious effects on galaxy scaling relations.

\section{Conclusions}\label{conclude}

hCOSMOS is a dense redshift survey targeting intermediate redshift galaxies in the COSMOS field.  
A subset of the survey, hCOS20.6, is uniformly $>90$\% complete  across the central $0.89$~deg$^2$ of the COSMOS field to a limiting $r=20.6$. hCOSMOS spectra cover the $3700-9100$~\AA\ wavelength range at a resolution of $R\sim1500$. hCOSMOS20.6 includes 1968 galaxies covering the redshift range $0.01\lesssim z \lesssim 0.6$ with a median redshift of $\widetilde z=0.27$.

The completeness of the survey and the broad wavelength coverage of the spectra make hCOS20.6  unique.
We measure the D$_n4000$ index, an indicator of quiescence and a proxy for galaxy age, for all hCOS20.6 galaxies. The completeness of the survey has already  provided a platform for assessing the environment of compact quiescent galaxies at intermediate redshifts \citep{Damjanov2015b}.  

In preparation for much larger, deeper future surveys, we demonstrate some uses of hCOS20.6.
Comparison of the D$_n4000$ distribution for the magnitude limited sample with their rest-frame $UVJ$ colors confirms that D$_n4000=1.5$ efficiently separates star-forming and quiescent populations. In contrast with the rest-frame colors, the D$_n4000$ index is insensitive to reddening and does not require $K-$correction. Although the small area of hCOS20.6 survey makes cosmic variance a serious issue, comparison with a larger area complete redshift survey \citep[SHELS F2 field,][]{Geller2014} shows that, averaged over the hCOS20.6 volume, the fraction of star-forming and quiescent systems is insensitive to cosmic variance. Averaged over the entire field, the hCOSMOS survey probes a range of environments such that the scaling relations we explore also appear to be insensitive to cosmic variance.
 
The COSMOS regions (and future regions imaged by, for example, EUCLID) enable the combination of spectral parameters like  D$_n4000$ indices with photometric parameters as probes of galaxy evolution. We demonstrate, for example,  that compact galaxies in hCOS20.6 are generally quiescent (with D$_n4000\gtrsim1.5$). Furthermore, there is a trend in compactness \citep[defined as in][]{Barro2013} with D$_n4000$ index as a proxy of galaxy age:  the most compact systems contain the oldest stellar population (D$_n4000>1.8$). This trend largely reflects the larger stellar masses of galaxies with high D$_n4000$ indices.   
 
Spectroscopically determined central velocity dispersions in the COSMOS region have not been available previously.
We measure velocity dispersions for 762 objects in the quiescent galaxy population of hCOS20.6 ($85\%$ of the quiescent sample). Together with galaxy size and stellar mass, these velocity dispersion measurements probe the stellar mass fundamental plane of quiescent systems at $0.1<z<0.6$ as a function of galaxy compactness \citep{Zahid2016a}. 

Here we use the measured velocity dispersions to test a photometric velocity dispersion proxy \citep{Bezanson2011, Bezanson2012, Belli2014} at these intermediate redshifts. Differences between the measured and predicted velocity dispersions depend on the spectroscopic properties of galaxies (e.g, the measured velocity dispersion itself and D$_n4000$). These results underscore the importance of direct spectroscopic measurements for understanding the connections between galaxies and their dark matter halos \citep[e.g.,][]{Wake2012, Belli2014, Bogdan2015, Schechter2016, Zahid2016b, Zahid2017, Belli2017}. 
 
In combination with structural properties based on high-resolution HST imaging and stellar mass estimates from broad-band photometry, hCOSMOS provide unique probes of dynamical, stellar population and environmental properties of intermediate-redshift galaxies.  The hCOSMOS survey demonstrates the power of combining a complete magnitude limited spectroscopic survey with high-resolution imaging. Larger area surveys reaching to greater depth promise increasingly powerful fundamental constraints on the evolution of the baryonic properties of galaxies. Spectroscopic surveys that provide central velocity dispersions may be a route to connecting this evolution to the underlying dark matter halos.  

\begin{deluxetable*}{ccccccccc}
\tabletypesize{\small}
\tablecaption{Photometric and Spectroscopic Properties of the hCOSMOS Galaxy Sample\tablenotemark{a}\label{table2}}
\tablewidth{0in}
\tablehead{\colhead{R.A.} & \colhead{Dec.} & \colhead{Source\tablenotemark{b}} & \colhead{UltraVISTA\tablenotemark{c}} & \colhead{$z$} & \colhead{D$_n4000$} & \colhead{$\sigma$} & \colhead{$\log(M/M_\sun)$} & \colhead{Comment\tablenotemark{d}}\\
\colhead{$[\arcdeg]$} & \colhead{$[\arcdeg]$} & \colhead{} & \colhead{} & \colhead{} & \colhead{}  & \colhead{[km s$^{-1}$]} & \colhead{ } & \colhead{}
}
\colnumbers 
\startdata
149.934367 & 2.199647 & hectospec & 119899 & 0.30178$\pm$0.00012 & 1.35$\pm$0.03 & -999$\pm$-999 & 10.22$^{+0.07}_{-0.06}$ & hCOS20.6 \\
149.835233 & 2.275360 & hectospec & 165304 & 0.21860$\pm$0.00009 & 2.03$\pm$0.09 & 136$\pm$25 & 10.41$^{+0.06}_{-0.06}$ & hCOS20.6 \\
150.127546 & 2.416870 & hectospec & 137878 & 0.25005$\pm$0.00006 & 1.16$\pm$0.08 & -999$\pm$-999 & 9.67$^{+0.13}_{-0.36}$ & hCOS20.6 \\
149.883800 & 1.827546 & hectospec & 59119 & 0.12406$\pm$0.00015 & 1.47$\pm$0.06 & -999$\pm$-999 & 9.70$^{+0.11}_{-0.12}$ & hCOS20.6 \\
150.351992 & 1.802384 & hectospec & 29063 & 0.29967$\pm$0.00013 & 1.21$\pm$0.05 & -999$\pm$-999 & 9.83$^{+0.12}_{-0.11}$ & hCOS20.6 \\
149.799300 & 2.169410 & hectospec & 156250 & 0.26188$\pm$0.00014 & 1.71$\pm$0.11 & 148$\pm$43 & 10.42$^{+0.07}_{-0.16}$ & hCOS20.6 \\
150.021588 & 2.314764 & hectospec & 129281 & 0.22071$\pm$0.00016 & 1.74$\pm$0.10 & -999$\pm$-999 & 10.11$^{+0.06}_{-0.16}$ & hCOS20.6\\ 
150.148058 & 1.784232 & hectospec & 28023 & 0.34671$\pm$0.00013 & 1.70$\pm$0.06 & 158$\pm$26 & 10.99$^{+0.05}_{-0.05}$ & hCOS20.6 \\
150.094888 & 2.297028 & hectospec & 127917 & 0.35973$\pm$0.00014 & 1.86$\pm$0.09 & 153$\pm$25 & 10.55$^{+0.14}_{-0.11}$ & hCOS20.6 \\
149.914554 & 1.783330 & hectospec & 27831 & 0.26608$\pm$0.00015 & 1.44$\pm$0.06 & -999$\pm$-999 & 10.61$^{+0.06}_{-0.07}$ & hCOS20.6 
\enddata
\tablecomments{
\tablenotetext{a}{This table is available in its entirety in a machine-readable form in the online journal. A portion is shown here for guidance regarding its form and content.}
\tablenotetext{b}{Instrument used for spectroscopic observations (Hectospec, SDSS, or BOSS)}
\tablenotetext{c}{Running sequence number in the \citet{Muzzin2013} photometric catalog}
\tablenotetext{d}{Comment: hCOS20.6 (subset of the survey that is $>90\%$ complete to $r=20.6$) or hCOSMOS ($r<21.3$ survey)}}
\end{deluxetable*}

\acknowledgements

We thank the anonymous reviewer for many useful comments including the suggestion to examine the relation between stellar mass and velocity dispersion. ID acknowledges the support of the Menzel Postdoctoral Fellowship during the data acquisition and analysis stage of the survey. At present ID is supported by the Canada Research Chair Program. HJZ gratefully acknowledges the generous support of the Clay Postdoctoral Fellowship. MJG is supported by the Smithsonian Institution.


\begin{thebibliography}{}
\expandafter\ifx\csname natexlab\endcsname\relax\def\natexlab#1{#1}\fi

\bibitem[{{Abraham} {et~al.}(2004){Abraham}, {Glazebrook}, {McCarthy},
  {Crampton}, {Murowinski}, {J{\o}rgensen}, {Roth}, {Hook}, {Savaglio}, {Chen},
  {Marzke}, \& {Carlberg}}]{Abraham2004}
{Abraham}, R.~G., {Glazebrook}, K., {McCarthy}, P.~J., {et~al.} 2004, \aj, 127,
  2455

\bibitem[{{Alam} {et~al.}(2015){Alam}, {Albareti}, {Allende Prieto}, {Anders},
  {Anderson}, {Anderton}, {Andrews}, {Armengaud}, {Aubourg}, {Bailey}, \&
  et~al.}]{Alam2015}
{Alam}, S., {Albareti}, F.~D., {Allende Prieto}, C., {et~al.} 2015, \apjs, 219,
  12

\bibitem[{{Arnouts} {et~al.}(1999){Arnouts}, {Cristiani}, {Moscardini},
  {Matarrese}, {Lucchin}, {Fontana}, \& {Giallongo}}]{Arnouts1999}
{Arnouts}, S., {Cristiani}, S., {Moscardini}, L., {et~al.} 1999, \mnras, 310,
  540

\bibitem[{{Baldry} {et~al.}(2010){Baldry}, {Robotham}, {Hill}, {Driver},
  {Liske}, {Norberg}, {Bamford}, {Hopkins}, {Loveday}, {Peacock}, {Cameron},
  {Croom}, {Cross}, {Doyle}, {Dye}, {Frenk}, {Jones}, {van Kampen}, {Kelvin},
  {Nichol}, {Parkinson}, {Popescu}, {Prescott}, {Sharp}, {Sutherland},
  {Thomas}, \& {Tuffs}}]{Baldry2010}
{Baldry}, I.~K., {Robotham}, A.~S.~G., {Hill}, D.~T., {et~al.} 2010, \mnras,
  404, 86

\bibitem[{{Balogh} {et~al.}(1999){Balogh}, {Morris}, {Yee}, {Carlberg}, \&
  {Ellingson}}]{Balogh1999}
{Balogh}, M.~L., {Morris}, S.~L., {Yee}, H.~K.~C., {Carlberg}, R.~G., \&
  {Ellingson}, E. 1999, \apj, 527, 54

\bibitem[{{Barro} {et~al.}(2013){Barro}, {Faber}, {P{\'e}rez-Gonz{\'a}lez},
  {Koo}, {Williams}, {Kocevski}, {Trump}, {Mozena}, {McGrath}, {van der Wel},
  {Wuyts}, {Bell}, {Croton}, {Ceverino}, {Dekel}, {Ashby}, {Cheung},
  {Ferguson}, {Fontana}, {Fang}, {Giavalisco}, {Grogin}, {Guo}, {Hathi},
  {Hopkins}, {Huang}, {Koekemoer}, {Kartaltepe}, {Lee}, {Newman}, {Porter},
  {Primack}, {Ryan}, {Rosario}, {Somerville}, {Salvato}, \& {Hsu}}]{Barro2013}
{Barro}, G., {Faber}, S.~M., {P{\'e}rez-Gonz{\'a}lez}, P.~G., {et~al.} 2013,
  \apj, 765, 104

\bibitem[{{Belli} {et~al.}(2014){Belli}, {Newman}, \& {Ellis}}]{Belli2014}
{Belli}, S., {Newman}, A.~B., \& {Ellis}, R.~S. 2014, \apj, 783, 117

\bibitem[{{Belli} {et~al.}(2017){Belli}, {Newman}, \& {Ellis}}]{Belli2017}
---. 2017, \apj, 834, 18

\bibitem[{{Bertin} {et~al.}(2002){Bertin}, {Ciotti}, \& {Del
  Principe}}]{Bertin2002}
{Bertin}, G., {Ciotti}, L., \& {Del Principe}, M. 2002, \aap, 386, 149

\bibitem[{{Bezanson} {et~al.}(2012){Bezanson}, {van Dokkum}, \&
  {Franx}}]{Bezanson2012}
{Bezanson}, R., {van Dokkum}, P., \& {Franx}, M. 2012, \apj, 760, 62

\bibitem[{{Bezanson} {et~al.}(2011){Bezanson}, {van Dokkum}, {Franx},
  {Brammer}, {Brinchmann}, {Kriek}, {Labb{\'e}}, {Quadri}, {Rix}, {van de
  Sande}, {Whitaker}, \& {Williams}}]{Bezanson2011}
{Bezanson}, R., {van Dokkum}, P.~G., {Franx}, M., {et~al.} 2011, \apjl, 737,
  L31

\bibitem[{{Bogd{\'a}n} \& {Goulding}(2015)}]{Bogdan2015}
{Bogd{\'a}n}, {\'A}., \& {Goulding}, A.~D. 2015, \apj, 800, 124

\bibitem[{{Bruzual} \& {Charlot}(2003)}]{Bruzual2003}
{Bruzual}, G., \& {Charlot}, S. 2003, \mnras, 344, 1000

\bibitem[{{Calzetti} {et~al.}(2000){Calzetti}, {Armus}, {Bohlin}, {Kinney},
  {Koornneef}, \& {Storchi-Bergmann}}]{Calzetti2000}
{Calzetti}, D., {Armus}, L., {Bohlin}, R.~C., {et~al.} 2000, \apj, 533, 682

\bibitem[{{Capak} {et~al.}(2007){Capak}, {Aussel}, {Ajiki}, {McCracken},
  {Mobasher}, {Scoville}, {Shopbell}, {Taniguchi}, {Thompson}, {Tribiano},
  {Sasaki}, {Blain}, {Brusa}, {Carilli}, {Comastri}, {Carollo}, {Cassata},
  {Colbert}, {Ellis}, {Elvis}, {Giavalisco}, {Green}, {Guzzo}, {Hasinger},
  {Ilbert}, {Impey}, {Jahnke}, {Kartaltepe}, {Kneib}, {Koda}, {Koekemoer},
  {Komiyama}, {Leauthaud}, {Le Fevre}, {Lilly}, {Liu}, {Massey}, {Miyazaki},
  {Murayama}, {Nagao}, {Peacock}, {Pickles}, {Porciani}, {Renzini}, {Rhodes},
  {Rich}, {Salvato}, {Sanders}, {Scarlata}, {Schiminovich}, {Schinnerer},
  {Scodeggio}, {Sheth}, {Shioya}, {Tasca}, {Taylor}, {Yan}, \&
  {Zamorani}}]{Capak2007}
{Capak}, P., {Aussel}, H., {Ajiki}, M., {et~al.} 2007, \apjs, 172, 99

\bibitem[{{Cappellari} {et~al.}(2006){Cappellari}, {Bacon}, {Bureau}, {Damen},
  {Davies}, {de Zeeuw}, {Emsellem}, {Falc{\'o}n-Barroso}, {Krajnovi{\'c}},
  {Kuntschner}, {McDermid}, {Peletier}, {Sarzi}, {van den Bosch}, \& {van de
  Ven}}]{Cappellari2006}
{Cappellari}, M., {Bacon}, R., {Bureau}, M., {et~al.} 2006, \mnras, 366, 1126

\bibitem[{{Chabrier}(2003)}]{Chabrier2003}
{Chabrier}, G. 2003, \pasp, 115, 763

\bibitem[{{Coil} {et~al.}(2011){Coil}, {Blanton}, {Burles}, {Cool},
  {Eisenstein}, {Moustakas}, {Wong}, {Zhu}, {Aird}, {Bernstein}, {Bolton}, \&
  {Hogg}}]{Coil2011}
{Coil}, A.~L., {Blanton}, M.~R., {Burles}, S.~M., {et~al.} 2011, \apj, 741, 8

\bibitem[{{Colless} {et~al.}(2001){Colless}, {Dalton}, {Maddox}, {Sutherland},
  {Norberg}, {Cole}, {Bland-Hawthorn}, {Bridges}, {Cannon}, {Collins}, {Couch},
  {Cross}, {Deeley}, {De Propris}, {Driver}, {Efstathiou}, {Ellis}, {Frenk},
  {Glazebrook}, {Jackson}, {Lahav}, {Lewis}, {Lumsden}, {Madgwick}, {Peacock},
  {Peterson}, {Price}, {Seaborne}, \& {Taylor}}]{Colless2001}
{Colless}, M., {Dalton}, G., {Maddox}, S., {et~al.} 2001, \mnras, 328, 1039

\bibitem[{{da Costa}(1998)}]{daCosta1998}
{da Costa}, L. 1998, ArXiv Astrophysics e-prints, astro-ph/9812258

\bibitem[{{Damjanov} {et~al.}(2015{\natexlab{a}}){Damjanov}, {Geller}, {Zahid},
  \& {Hwang}}]{Damjanov2015a}
{Damjanov}, I., {Geller}, M.~J., {Zahid}, H.~J., \& {Hwang}, H.~S.
  2015{\natexlab{a}}, \apj, 806, 158

\bibitem[{{Damjanov} {et~al.}(2015{\natexlab{b}}){Damjanov}, {Zahid}, {Geller},
  \& {Hwang}}]{Damjanov2015b}
{Damjanov}, I., {Zahid}, H.~J., {Geller}, M.~J., \& {Hwang}, H.~S.
  2015{\natexlab{b}}, \apj, 815, 104

\bibitem[{{Damjanov} {et~al.}(2011){Damjanov}, {Abraham}, {Glazebrook},
  {McCarthy}, {Caris}, {Carlberg}, {Chen}, {Crampton}, {Green}, {J{\o}rgensen},
  {Juneau}, {Le Borgne}, {Marzke}, {Mentuch}, {Murowinski}, {Roth}, {Savaglio},
  \& {Yan}}]{Damjanov2011}
{Damjanov}, I., {Abraham}, R.~G., {Glazebrook}, K., {et~al.} 2011, \apjl, 739,
  L44

\bibitem[{{Davis} {et~al.}(2003){Davis}, {Faber}, {Newman}, {Phillips},
  {Ellis}, {Steidel}, {Conselice}, {Coil}, {Finkbeiner}, {Koo}, {Guhathakurta},
  {Weiner}, {Schiavon}, {Willmer}, {Kaiser}, {Luppino}, {Wirth}, {Connolly},
  {Eisenhardt}, {Cooper}, \& {Gerke}}]{Davis2003}
{Davis}, M., {Faber}, S.~M., {Newman}, J., {et~al.} 2003, in \procspie, Vol.
  4834, Discoveries and Research Prospects from 6- to 10-Meter-Class Telescopes
  II, ed. P.~{Guhathakurta}, 161--172

\bibitem[{{Delaye} {et~al.}(2014){Delaye}, {Huertas-Company}, {Mei}, {Lidman},
  {Licitra}, {Newman}, {Raichoor}, {Shankar}, {Barrientos}, {Bernardi},
  {Cerulo}, {Couch}, {Demarco}, {Mu{\~n}oz}, {S{\'a}nchez-Janssen}, \&
  {Tanaka}}]{Delaye2014}
{Delaye}, L., {Huertas-Company}, M., {Mei}, S., {et~al.} 2014, \mnras, 441, 203

\bibitem[{{Fabricant} {et~al.}(2013){Fabricant}, {Chilingarian}, {Hwang},
  {Kurtz}, {Geller}, {Del'Antonio}, \& {Rines}}]{Fabricant2013}
{Fabricant}, D., {Chilingarian}, I., {Hwang}, H.~S., {et~al.} 2013, \pasp, 125,
  1362

\bibitem[{{Fabricant} {et~al.}(2005){Fabricant}, {Fata}, {Roll}, {Hertz},
  {Caldwell}, {Gauron}, {Geary}, {McLeod}, {Szentgyorgyi}, {Zajac}, {Kurtz},
  {Barberis}, {Bergner}, {Brown}, {Conroy}, {Eng}, {Geller}, {Goddard},
  {Honsa}, {Mueller}, {Mink}, {Ordway}, {Tokarz}, {Woods}, {Wyatt}, {Epps}, \&
  {Dell'Antonio}}]{Fabricant2005}
{Fabricant}, D., {Fata}, R., {Roll}, J., {et~al.} 2005, \pasp, 117, 1411

\bibitem[{{Fabricant} {et~al.}(1998){Fabricant}, {Hertz}, {Szentgyorgyi},
  {Fata}, {Roll}, \& {Zajac}}]{Fabricant1998}
{Fabricant}, D.~G., {Hertz}, E.~N., {Szentgyorgyi}, A.~H., {et~al.} 1998, in
  \procspie, Vol. 3355, Optical Astronomical Instrumentation, ed.
  S.~{D'Odorico}, 285--296

\bibitem[{{Fabricant} {et~al.}(2008){Fabricant}, {Kurtz}, {Geller}, {Caldwell},
  {Woods}, \& {Dell'Antonio}}]{Fabricant2008}
{Fabricant}, D.~G., {Kurtz}, M.~J., {Geller}, M.~J., {et~al.} 2008, \pasp, 120,
  1222

\bibitem[{{Geller} \& {Huchra}(1989)}]{Geller1989}
{Geller}, M.~J., \& {Huchra}, J.~P. 1989, Science, 246, 897

\bibitem[{{Geller} {et~al.}(2016){Geller}, {Hwang}, {Dell'Antonio}, {Zahid},
  {Kurtz}, \& {Fabricant}}]{Geller2016}
{Geller}, M.~J., {Hwang}, H.~S., {Dell'Antonio}, I.~P., {et~al.} 2016, \apjs,
  224, 11

\bibitem[{{Geller} {et~al.}(2014){Geller}, {Hwang}, {Fabricant}, {Kurtz},
  {Dell'Antonio}, \& {Zahid}}]{Geller2014}
{Geller}, M.~J., {Hwang}, H.~S., {Fabricant}, D.~G., {et~al.} 2014, \apjs, 213,
  35

\bibitem[{{Guzzo} {et~al.}(2014){Guzzo}, {Scodeggio}, {Garilli}, {Granett},
  {Fritz}, {Abbas}, {Adami}, {Arnouts}, {Bel}, {Bolzonella}, {Bottini},
  {Branchini}, {Cappi}, {Coupon}, {Cucciati}, {Davidzon}, {De Lucia}, {de la
  Torre}, {Franzetti}, {Fumana}, {Hudelot}, {Ilbert}, {Iovino}, {Krywult}, {Le
  Brun}, {Le F{\`e}vre}, {Maccagni}, {Ma{\l}ek}, {Marulli}, {McCracken},
  {Paioro}, {Peacock}, {Polletta}, {Pollo}, {Schlagenhaufer}, {Tasca},
  {Tojeiro}, {Vergani}, {Zamorani}, {Zanichelli}, {Burden}, {Di Porto},
  {Marchetti}, {Marinoni}, {Mellier}, {Moscardini}, {Nichol}, {Percival},
  {Phleps}, \& {Wolk}}]{Guzzo2014}
{Guzzo}, L., {Scodeggio}, M., {Garilli}, B., {et~al.} 2014, \aap, 566, A108

\bibitem[{{Haines} {et~al.}(2017){Haines}, {Iovino}, {Krywult}, {Guzzo},
  {Davidzon}, {Bolzonella}, {Garilli}, {Scodeggio}, {Granett}, {de la Torre},
  {De Lucia}, {Abbas}, {Adami}, {Arnouts}, {Bottini}, {Cappi}, {Cucciati},
  {Franzetti}, {Fritz}, {Gargiulo}, {Le Brun}, {Le F{\`e}vre}, {Maccagni},
  {Ma{\l}ek}, {Marulli}, {Moutard}, {Polletta}, {Pollo}, {Tasca}, {Tojeiro},
  {Vergani}, {Zanichelli}, {Zamorani}, {Bel}, {Branchini}, {Coupon}, {Ilbert},
  {Moscardini}, {Peacock}, \& {Siudek}}]{Haines2017}
{Haines}, C.~P., {Iovino}, A., {Krywult}, J., {et~al.} 2017, \aap, 605, A4

\bibitem[{{Huertas-Company} {et~al.}(2013){Huertas-Company}, {Shankar}, {Mei},
  {Bernardi}, {Aguerri}, {Meert}, \& {Vikram}}]{HuertasCompany2013}
{Huertas-Company}, M., {Shankar}, F., {Mei}, S., {et~al.} 2013, \apj, 779, 29

\bibitem[{{Ilbert} {et~al.}(2006){Ilbert}, {Arnouts}, {McCracken},
  {Bolzonella}, {Bertin}, {Le F{\`e}vre}, {Mellier}, {Zamorani}, {Pell{\`o}},
  {Iovino}, {Tresse}, {Le Brun}, {Bottini}, {Garilli}, {Maccagni}, {Picat},
  {Scaramella}, {Scodeggio}, {Vettolani}, {Zanichelli}, {Adami}, {Bardelli},
  {Cappi}, {Charlot}, {Ciliegi}, {Contini}, {Cucciati}, {Foucaud}, {Franzetti},
  {Gavignaud}, {Guzzo}, {Marano}, {Marinoni}, {Mazure}, {Meneux}, {Merighi},
  {Paltani}, {Pollo}, {Pozzetti}, {Radovich}, {Zucca}, {Bondi}, {Bongiorno},
  {Busarello}, {de La Torre}, {Gregorini}, {Lamareille}, {Mathez}, {Merluzzi},
  {Ripepi}, {Rizzo}, \& {Vergani}}]{Ilbert2006}
{Ilbert}, O., {Arnouts}, S., {McCracken}, H.~J., {et~al.} 2006, \aap, 457, 841

\bibitem[{{Jones} {et~al.}(2009){Jones}, {Read}, {Saunders}, {Colless},
  {Jarrett}, {Parker}, {Fairall}, {Mauch}, {Sadler}, {Watson}, {Burton},
  {Campbell}, {Cass}, {Croom}, {Dawe}, {Fiegert}, {Frankcombe}, {Hartley},
  {Huchra}, {James}, {Kirby}, {Lahav}, {Lucey}, {Mamon}, {Moore}, {Peterson},
  {Prior}, {Proust}, {Russell}, {Safouris}, {Wakamatsu}, {Westra}, \&
  {Williams}}]{Jones2009}
{Jones}, D.~H., {Read}, M.~A., {Saunders}, W., {et~al.} 2009, \mnras, 399, 683

\bibitem[{{Jorgensen} {et~al.}(1995){Jorgensen}, {Franx}, \&
  {Kjaergaard}}]{Jorgensen1995}
{Jorgensen}, I., {Franx}, M., \& {Kjaergaard}, P. 1995, \mnras, 276, 1341

\bibitem[{{Kauffmann} {et~al.}(2003){Kauffmann}, {Heckman}, {White}, {Charlot},
  {Tremonti}, {Brinchmann}, {Bruzual}, {Peng}, {Seibert}, {Bernardi},
  {Blanton}, {Brinkmann}, {Castander}, {Cs{\'a}bai}, {Fukugita}, {Ivezic},
  {Munn}, {Nichol}, {Padmanabhan}, {Thakar}, {Weinberg}, \&
  {York}}]{Kauffmann2003}
{Kauffmann}, G., {Heckman}, T.~M., {White}, S.~D.~M., {et~al.} 2003, \mnras,
  341, 33

\bibitem[{{Kochanek} {et~al.}(2012){Kochanek}, {Eisenstein}, {Cool},
  {Caldwell}, {Assef}, {Jannuzi}, {Jones}, {Murray}, {Forman}, {Dey}, {Brown},
  {Eisenhardt}, {Gonzalez}, {Green}, \& {Stern}}]{Kochanek2012}
{Kochanek}, C.~S., {Eisenstein}, D.~J., {Cool}, R.~J., {et~al.} 2012, \apjs,
  200, 8

\bibitem[{{Koleva} {et~al.}(2009){Koleva}, {Prugniel}, {Bouchard}, \&
  {Wu}}]{Koleva2009}
{Koleva}, M., {Prugniel}, P., {Bouchard}, A., \& {Wu}, Y. 2009, \aap, 501, 1269

\bibitem[{{Kurtz} \& {Mink}(1998)}]{Kurtz1998}
{Kurtz}, M.~J., \& {Mink}, D.~J. 1998, \pasp, 110, 934

\bibitem[{{Lange} {et~al.}(2015){Lange}, {Driver}, {Robotham}, {Kelvin},
  {Graham}, {Alpaslan}, {Andrews}, {Baldry}, {Bamford}, {Bland-Hawthorn},
  {Brough}, {Cluver}, {Conselice}, {Davies}, {Haeussler}, {Konstantopoulos},
  {Loveday}, {Moffett}, {Norberg}, {Phillipps}, {Taylor},
  {L{\'o}pez-S{\'a}nchez}, \& {Wilkins}}]{Lange2015}
{Lange}, R., {Driver}, S.~P., {Robotham}, A.~S.~G., {et~al.} 2015, \mnras, 447,
  2603

\bibitem[{{Lange} {et~al.}(2016){Lange}, {Moffett}, {Driver}, {Robotham},
  {Lagos}, {Kelvin}, {Conselice}, {Margalef-Bentabol}, {Alpaslan}, {Baldry},
  {Bland-Hawthorn}, {Bremer}, {Brough}, {Cluver}, {Colless}, {Davies},
  {H{\"a}u{\ss}ler}, {Holwerda}, {Hopkins}, {Kafle}, {Kennedy}, {Liske},
  {Phillipps}, {Popescu}, {Taylor}, {Tuffs}, {van Kampen}, \&
  {Wright}}]{Lange2016}
{Lange}, R., {Moffett}, A.~J., {Driver}, S.~P., {et~al.} 2016, \mnras, 462,
  1470

\bibitem[{{Le Borgne} {et~al.}(2004){Le Borgne}, {Rocca-Volmerange},
  {Prugniel}, {Lan{\c c}on}, {Fioc}, \& {Soubiran}}]{LeBorgne2004}
{Le Borgne}, D., {Rocca-Volmerange}, B., {Prugniel}, P., {et~al.} 2004, \aap,
  425, 881

\bibitem[{{Le F{\`e}vre} {et~al.}(2005){Le F{\`e}vre}, {Vettolani}, {Garilli},
  {Tresse}, {Bottini}, {Le Brun}, {Maccagni}, {Picat}, {Scaramella},
  {Scodeggio}, {Zanichelli}, {Adami}, {Arnaboldi}, {Arnouts}, {Bardelli},
  {Bolzonella}, {Cappi}, {Charlot}, {Ciliegi}, {Contini}, {Foucaud},
  {Franzetti}, {Gavignaud}, {Guzzo}, {Ilbert}, {Iovino}, {McCracken}, {Marano},
  {Marinoni}, {Mathez}, {Mazure}, {Meneux}, {Merighi}, {Paltani}, {Pell{\`o}},
  {Pollo}, {Pozzetti}, {Radovich}, {Zamorani}, {Zucca}, {Bondi}, {Bongiorno},
  {Busarello}, {Lamareille}, {Mellier}, {Merluzzi}, {Ripepi}, \&
  {Rizzo}}]{LeFevre2005}
{Le F{\`e}vre}, O., {Vettolani}, G., {Garilli}, B., {et~al.} 2005, \aap, 439,
  845

\bibitem[{{Le F{\`e}vre} {et~al.}(2015){Le F{\`e}vre}, {Tasca}, {Cassata},
  {Garilli}, {Le Brun}, {Maccagni}, {Pentericci}, {Thomas}, {Vanzella},
  {Zamorani}, {Zucca}, {Amorin}, {Bardelli}, {Capak}, {Cassar{\`a}},
  {Castellano}, {Cimatti}, {Cuby}, {Cucciati}, {de la Torre}, {Durkalec},
  {Fontana}, {Giavalisco}, {Grazian}, {Hathi}, {Ilbert}, {Lemaux}, {Moreau},
  {Paltani}, {Ribeiro}, {Salvato}, {Schaerer}, {Scodeggio}, {Sommariva},
  {Talia}, {Taniguchi}, {Tresse}, {Vergani}, {Wang}, {Charlot}, {Contini},
  {Fotopoulou}, {L{\'o}pez-Sanjuan}, {Mellier}, \& {Scoville}}]{LeFevre2015}
{Le F{\`e}vre}, O., {Tasca}, L.~A.~M., {Cassata}, P., {et~al.} 2015, \aap, 576,
  A79

\bibitem[{{Lilly} {et~al.}(2007){Lilly}, {Le F{\`e}vre}, {Renzini}, {Zamorani},
  {Scodeggio}, {Contini}, {Carollo}, {Hasinger}, {Kneib}, {Iovino}, {Le Brun},
  {Maier}, {Mainieri}, {Mignoli}, {Silverman}, {Tasca}, {Bolzonella},
  {Bongiorno}, {Bottini}, {Capak}, {Caputi}, {Cimatti}, {Cucciati}, {Daddi},
  {Feldmann}, {Franzetti}, {Garilli}, {Guzzo}, {Ilbert}, {Kampczyk}, {Kovac},
  {Lamareille}, {Leauthaud}, {Le Borgne}, {McCracken}, {Marinoni}, {Pello},
  {Ricciardelli}, {Scarlata}, {Vergani}, {Sanders}, {Schinnerer}, {Scoville},
  {Taniguchi}, {Arnouts}, {Aussel}, {Bardelli}, {Brusa}, {Cappi}, {Ciliegi},
  {Finoguenov}, {Foucaud}, {Franceschini}, {Halliday}, {Impey}, {Knobel},
  {Koekemoer}, {Kurk}, {Maccagni}, {Maddox}, {Marano}, {Marconi}, {Meneux},
  {Mobasher}, {Moreau}, {Peacock}, {Porciani}, {Pozzetti}, {Scaramella},
  {Schiminovich}, {Shopbell}, {Smail}, {Thompson}, {Tresse}, {Vettolani},
  {Zanichelli}, \& {Zucca}}]{Lilly2007}
{Lilly}, S.~J., {Le F{\`e}vre}, O., {Renzini}, A., {et~al.} 2007, \apjs, 172,
  70

\bibitem[{{Lilly} {et~al.}(2009){Lilly}, {Le Brun}, {Maier}, {Mainieri},
  {Mignoli}, {Scodeggio}, {Zamorani}, {Carollo}, {Contini}, {Kneib}, {Le
  F{\`e}vre}, {Renzini}, {Bardelli}, {Bolzonella}, {Bongiorno}, {Caputi},
  {Coppa}, {Cucciati}, {de la Torre}, {de Ravel}, {Franzetti}, {Garilli},
  {Iovino}, {Kampczyk}, {Kovac}, {Knobel}, {Lamareille}, {Le Borgne}, {Pello},
  {Peng}, {P{\'e}rez-Montero}, {Ricciardelli}, {Silverman}, {Tanaka}, {Tasca},
  {Tresse}, {Vergani}, {Zucca}, {Ilbert}, {Salvato}, {Oesch}, {Abbas},
  {Bottini}, {Capak}, {Cappi}, {Cassata}, {Cimatti}, {Elvis}, {Fumana},
  {Guzzo}, {Hasinger}, {Koekemoer}, {Leauthaud}, {Maccagni}, {Marinoni},
  {McCracken}, {Memeo}, {Meneux}, {Porciani}, {Pozzetti}, {Sanders},
  {Scaramella}, {Scarlata}, {Scoville}, {Shopbell}, \& {Taniguchi}}]{Lilly2009}
{Lilly}, S.~J., {Le Brun}, V., {Maier}, C., {et~al.} 2009, \apjs, 184, 218

\bibitem[{{Loveday} {et~al.}(1992){Loveday}, {Efstathiou}, {Peterson}, \&
  {Maddox}}]{Loveday1992}
{Loveday}, J., {Efstathiou}, G., {Peterson}, B.~A., \& {Maddox}, S.~J. 1992,
  \apjl, 400, L43

\bibitem[{{Maltby} {et~al.}(2010){Maltby}, {Arag{\'o}n-Salamanca}, {Gray},
  {Barden}, {H{\"a}u{\ss}ler}, {Wolf}, {Peng}, {Jahnke}, {McIntosh},
  {B{\"o}hm}, \& {van Kampen}}]{Maltby2010}
{Maltby}, D.~T., {Arag{\'o}n-Salamanca}, A., {Gray}, M.~E., {et~al.} 2010,
  \mnras, 402, 282

\bibitem[{{Martin} {et~al.}(2005){Martin}, {Fanson}, {Schiminovich},
  {Morrissey}, {Friedman}, {Barlow}, {Conrow}, {Grange}, {Jelinsky},
  {Milliard}, {Siegmund}, {Bianchi}, {Byun}, {Donas}, {Forster}, {Heckman},
  {Lee}, {Madore}, {Malina}, {Neff}, {Rich}, {Small}, {Surber}, {Szalay},
  {Welsh}, \& {Wyder}}]{Martin2005}
{Martin}, D.~C., {Fanson}, J., {Schiminovich}, D., {et~al.} 2005, \apjl, 619,
  L1

\bibitem[{{McCracken} {et~al.}(2012){McCracken}, {Milvang-Jensen}, {Dunlop},
  {Franx}, {Fynbo}, {Le F{\`e}vre}, {Holt}, {Caputi}, {Goranova}, {Buitrago},
  {Emerson}, {Freudling}, {Hudelot}, {L{\'o}pez-Sanjuan}, {Magnard}, {Mellier},
  {M{\o}ller}, {Nilsson}, {Sutherland}, {Tasca}, \& {Zabl}}]{McCracken2012}
{McCracken}, H.~J., {Milvang-Jensen}, B., {Dunlop}, J., {et~al.} 2012, \aap,
  544, A156

\bibitem[{{Mehlert} {et~al.}(2003){Mehlert}, {Thomas}, {Saglia}, {Bender}, \&
  {Wegner}}]{Mehlert2003}
{Mehlert}, D., {Thomas}, D., {Saglia}, R.~P., {Bender}, R., \& {Wegner}, G.
  2003, \aap, 407, 423

\bibitem[{{Moresco} {et~al.}(2013){Moresco}, {Pozzetti}, {Cimatti}, {Zamorani},
  {Bolzonella}, {Lamareille}, {Mignoli}, {Zucca}, {Lilly}, {Carollo},
  {Contini}, {Kneib}, {Le F{\`e}vre}, {Mainieri}, {Renzini}, {Scodeggio},
  {Bardelli}, {Bongiorno}, {Caputi}, {Cucciati}, {de la Torre}, {de Ravel},
  {Franzetti}, {Garilli}, {Iovino}, {Kampczyk}, {Knobel}, {Kova{\v c}}, {Le
  Borgne}, {Le Brun}, {Maier}, {Pell{\'o}}, {Peng}, {Perez-Montero},
  {Presotto}, {Silverman}, {Tanaka}, {Tasca}, {Tresse}, {Vergani}, {Barnes},
  {Bordoloi}, {Cappi}, {Diener}, {Koekemoer}, {Le Floc'h}, {L{\'o}pez-Sanjuan},
  {McCracken}, {Nair}, {Oesch}, {Scarlata}, {Scoville}, \&
  {Welikala}}]{Moresco2013}
{Moresco}, M., {Pozzetti}, L., {Cimatti}, A., {et~al.} 2013, \aap, 558, A61

\bibitem[{{Moustakas} {et~al.}(2013){Moustakas}, {Coil}, {Aird}, {Blanton},
  {Cool}, {Eisenstein}, {Mendez}, {Wong}, {Zhu}, \& {Arnouts}}]{Moustakas2013}
{Moustakas}, J., {Coil}, A.~L., {Aird}, J., {et~al.} 2013, \apj, 767, 50

\bibitem[{{Muzzin} {et~al.}(2013){Muzzin}, {Marchesini}, {Stefanon}, {Franx},
  {Milvang-Jensen}, {Dunlop}, {Fynbo}, {Brammer}, {Labb{\'e}}, \& {van
  Dokkum}}]{Muzzin2013}
{Muzzin}, A., {Marchesini}, D., {Stefanon}, M., {et~al.} 2013, \apjs, 206, 8

\bibitem[{{Newman} {et~al.}(2012){Newman}, {Ellis}, {Bundy}, \&
  {Treu}}]{Newman2012}
{Newman}, A.~B., {Ellis}, R.~S., {Bundy}, K., \& {Treu}, T. 2012, \apj, 746,
  162

\bibitem[{{Newman} {et~al.}(2013){Newman}, {Cooper}, {Davis}, {Faber}, {Coil},
  {Guhathakurta}, {Koo}, {Phillips}, {Conroy}, {Dutton}, {Finkbeiner}, {Gerke},
  {Rosario}, {Weiner}, {Willmer}, {Yan}, {Harker}, {Kassin}, {Konidaris},
  {Lai}, {Madgwick}, {Noeske}, {Wirth}, {Connolly}, {Kaiser}, {Kirby},
  {Lemaux}, {Lin}, {Lotz}, {Luppino}, {Marinoni}, {Matthews}, {Metevier}, \&
  {Schiavon}}]{Newman2013}
{Newman}, J.~A., {Cooper}, M.~C., {Davis}, M., {et~al.} 2013, \apjs, 208, 5

\bibitem[{{Noll} {et~al.}(2004){Noll}, {Mehlert}, {Appenzeller}, {Bender},
  {B{\"o}hm}, {Gabasch}, {Heidt}, {Hopp}, {J{\"a}ger}, {Seitz}, {Stahl},
  {Tapken}, \& {Ziegler}}]{Noll2004}
{Noll}, S., {Mehlert}, D., {Appenzeller}, I., {et~al.} 2004, \aap, 418, 885

\bibitem[{{Paulino-Afonso} {et~al.}(2017){Paulino-Afonso}, {Sobral},
  {Buitrago}, \& {Afonso}}]{PaulinoAfonso2017}
{Paulino-Afonso}, A., {Sobral}, D., {Buitrago}, F., \& {Afonso}, J. 2017,
  \mnras, 465, 2717

\bibitem[{{Poggianti} {et~al.}(2013){Poggianti}, {Calvi}, {Bindoni},
  {D'Onofrio}, {Moretti}, {Valentinuzzi}, {Fasano}, {Fritz}, {De Lucia},
  {Vulcani}, {Bettoni}, {Gullieuszik}, \& {Omizzolo}}]{Poggianti2013}
{Poggianti}, B.~M., {Calvi}, R., {Bindoni}, D., {et~al.} 2013, \apj, 762, 77

\bibitem[{{Roll} {et~al.}(1998){Roll}, {Fabricant}, \& {McLeod}}]{Roll1998}
{Roll}, J.~B., {Fabricant}, D.~G., \& {McLeod}, B.~A. 1998, in \procspie, Vol.
  3355, Optical Astronomical Instrumentation, ed. S.~{D'Odorico}, 324--332

\bibitem[{{S{\'a}nchez-Bl{\'a}zquez} {et~al.}(2006){S{\'a}nchez-Bl{\'a}zquez},
  {Peletier}, {Jim{\'e}nez-Vicente}, {Cardiel}, {Cenarro},
  {Falc{\'o}n-Barroso}, {Gorgas}, {Selam}, \& {Vazdekis}}]{SanchezBlazquez2006}
{S{\'a}nchez-Bl{\'a}zquez}, P., {Peletier}, R.~F., {Jim{\'e}nez-Vicente}, J.,
  {et~al.} 2006, \mnras, 371, 703

\bibitem[{{Sanders} {et~al.}(2007){Sanders}, {Salvato}, {Aussel}, {Ilbert},
  {Scoville}, {Surace}, {Frayer}, {Sheth}, {Helou}, {Brooke}, {Bhattacharya},
  {Yan}, {Kartaltepe}, {Barnes}, {Blain}, {Calzetti}, {Capak}, {Carilli},
  {Carollo}, {Comastri}, {Daddi}, {Ellis}, {Elvis}, {Fall}, {Franceschini},
  {Giavalisco}, {Hasinger}, {Impey}, {Koekemoer}, {Le F{\`e}vre}, {Lilly},
  {Liu}, {McCracken}, {Mobasher}, {Renzini}, {Rich}, {Schinnerer}, {Shopbell},
  {Taniguchi}, {Thompson}, {Urry}, \& {Williams}}]{Sanders2007}
{Sanders}, D.~B., {Salvato}, M., {Aussel}, H., {et~al.} 2007, \apjs, 172, 86

\bibitem[{{Sargent} {et~al.}(2007){Sargent}, {Carollo}, {Lilly}, {Scarlata},
  {Feldmann}, {Kampczyk}, {Koekemoer}, {Scoville}, {Kneib}, {Leauthaud},
  {Massey}, {Rhodes}, {Tasca}, {Capak}, {McCracken}, {Porciani}, {Renzini},
  {Taniguchi}, {Thompson}, \& {Sheth}}]{Sargent2007}
{Sargent}, M.~T., {Carollo}, C.~M., {Lilly}, S.~J., {et~al.} 2007, \apjs, 172,
  434

\bibitem[{{Scarlata} {et~al.}(2007){Scarlata}, {Carollo}, {Lilly}, {Sargent},
  {Feldmann}, {Kampczyk}, {Porciani}, {Koekemoer}, {Scoville}, {Kneib},
  {Leauthaud}, {Massey}, {Rhodes}, {Tasca}, {Capak}, {Maier}, {McCracken},
  {Mobasher}, {Renzini}, {Taniguchi}, {Thompson}, {Sheth}, {Ajiki}, {Aussel},
  {Murayama}, {Sanders}, {Sasaki}, {Shioya}, \& {Takahashi}}]{Scarlata2007}
{Scarlata}, C., {Carollo}, C.~M., {Lilly}, S., {et~al.} 2007, \apjs, 172, 406

\bibitem[{{Schechter}(2016)}]{Schechter2016}
{Schechter}, P.~L. 2016, in IAU Symposium, Vol. 317, The General Assembly of
  Galaxy Halos: Structure, Origin and Evolution, ed. A.~{Bragaglia},
  M.~{Arnaboldi}, M.~{Rejkuba}, \& D.~{Romano}, 35--38

\bibitem[{Scholz \& Stephens(1987)}]{Scholz1987}
Scholz, F.~W., \& Stephens, M.~A. 1987, Journal of the American Statistical
  Association, 82, 918

\bibitem[{{S{\'e}rsic}(1968)}]{Sersic1968}
{S{\'e}rsic}, J.~L. 1968, {Atlas de Galaxias Australes}

\bibitem[{{Shectman} {et~al.}(1996){Shectman}, {Landy}, {Oemler}, {Tucker},
  {Lin}, {Kirshner}, \& {Schechter}}]{Shectman1996}
{Shectman}, S.~A., {Landy}, S.~D., {Oemler}, A., {et~al.} 1996, \apj, 470, 172

\bibitem[{{Shen} {et~al.}(2003){Shen}, {Mo}, {White}, {Blanton}, {Kauffmann},
  {Voges}, {Brinkmann}, \& {Csabai}}]{Shen2003}
{Shen}, S., {Mo}, H.~J., {White}, S.~D.~M., {et~al.} 2003, \mnras, 343, 978

\bibitem[{{Silverman} {et~al.}(2015){Silverman}, {Kashino}, {Sanders},
  {Kartaltepe}, {Arimoto}, {Renzini}, {Rodighiero}, {Daddi}, {Zahid}, {Nagao},
  {Kewley}, {Lilly}, {Sugiyama}, {Baronchelli}, {Capak}, {Carollo}, {Chu},
  {Hasinger}, {Ilbert}, {Juneau}, {Kajisawa}, {Koekemoer}, {Kovac}, {Le
  F{\`e}vre}, {Masters}, {McCracken}, {Onodera}, {Schulze}, {Scoville},
  {Strazzullo}, \& {Taniguchi}}]{Silverman2015}
{Silverman}, J.~D., {Kashino}, D., {Sanders}, D., {et~al.} 2015, \apjs, 220, 12

\bibitem[{{Simard} {et~al.}(2002){Simard}, {Willmer}, {Vogt}, {Sarajedini},
  {Phillips}, {Weiner}, {Koo}, {Im}, {Illingworth}, \& {Faber}}]{Simard2002}
{Simard}, L., {Willmer}, C.~N.~A., {Vogt}, N.~P., {et~al.} 2002, \apjs, 142, 1

\bibitem[{{Steidel} {et~al.}(2003){Steidel}, {Adelberger}, {Shapley},
  {Pettini}, {Dickinson}, \& {Giavalisco}}]{Steidel2003}
{Steidel}, C.~C., {Adelberger}, K.~L., {Shapley}, A.~E., {et~al.} 2003, \apj,
  592, 728

\bibitem[{{Strauss} {et~al.}(2002){Strauss}, {Weinberg}, {Lupton}, {Narayanan},
  {Annis}, {Bernardi}, {Blanton}, {Burles}, {Connolly}, {Dalcanton}, {Doi},
  {Eisenstein}, {Frieman}, {Fukugita}, {Gunn}, {Ivezi{\'c}}, {Kent}, {Kim},
  {Knapp}, {Kron}, {Munn}, {Newberg}, {Nichol}, {Okamura}, {Quinn}, {Richmond},
  {Schlegel}, {Shimasaku}, {SubbaRao}, {Szalay}, {Vanden Berk}, {Vogeley},
  {Yanny}, {Yasuda}, {York}, \& {Zehavi}}]{Strauss2002}
{Strauss}, M.~A., {Weinberg}, D.~H., {Lupton}, R.~H., {et~al.} 2002, \aj, 124,
  1810

\bibitem[{{Sweet} {et~al.}(2017){Sweet}, {Sharp}, {Glazebrook}, {Rigaut},
  {Carrasco}, {Brodwin}, {Bayliss}, {Stalder}, {Abraham}, \&
  {McGregor}}]{Sweet2017}
{Sweet}, S.~M., {Sharp}, R., {Glazebrook}, K., {et~al.} 2017, \mnras, 464, 2910

\bibitem[{{Taniguchi} {et~al.}(2007){Taniguchi}, {Scoville}, {Murayama},
  {Sanders}, {Mobasher}, {Aussel}, {Capak}, {Ajiki}, {Miyazaki}, {Komiyama},
  {Shioya}, {Nagao}, {Sasaki}, {Koda}, {Carilli}, {Giavalisco}, {Guzzo},
  {Hasinger}, {Impey}, {LeFevre}, {Lilly}, {Renzini}, {Rich}, {Schinnerer},
  {Shopbell}, {Kaifu}, {Karoji}, {Arimoto}, {Okamura}, \&
  {Ohta}}]{Taniguchi2007}
{Taniguchi}, Y., {Scoville}, N., {Murayama}, T., {et~al.} 2007, \apjs, 172, 9

\bibitem[{{Taylor} {et~al.}(2010){Taylor}, {Franx}, {Glazebrook}, {Brinchmann},
  {van der Wel}, \& {van Dokkum}}]{Taylor2010}
{Taylor}, E.~N., {Franx}, M., {Glazebrook}, K., {et~al.} 2010, \apj, 720, 723

\bibitem[{{Tonry} \& {Davis}(1979)}]{Tonry1979}
{Tonry}, J., \& {Davis}, M. 1979, \aj, 84, 1511

\bibitem[{{Trujillo} {et~al.}(2007){Trujillo}, {Conselice}, {Bundy}, {Cooper},
  {Eisenhardt}, \& {Ellis}}]{Trujillo2007}
{Trujillo}, I., {Conselice}, C.~J., {Bundy}, K., {et~al.} 2007, \mnras, 382,
  109

\bibitem[{{Trujillo} {et~al.}(2004){Trujillo}, {Rudnick}, {Rix}, {Labb{\'e}},
  {Franx}, {Daddi}, {van Dokkum}, {F{\"o}rster Schreiber}, {Kuijken},
  {Moorwood}, {R{\"o}ttgering}, {van der Wel}, {van der Werf}, \& {van
  Starkenburg}}]{Trujillo2004}
{Trujillo}, I., {Rudnick}, G., {Rix}, H.-W., {et~al.} 2004, \apj, 604, 521

\bibitem[{{van der Wel} {et~al.}(2014){van der Wel}, {Franx}, {van Dokkum},
  {Skelton}, {Momcheva}, {Whitaker}, {Brammer}, {Bell}, {Rix}, {Wuyts},
  {Ferguson}, {Holden}, {Barro}, {Koekemoer}, {Chang}, {McGrath},
  {H{\"a}ussler}, {Dekel}, {Behroozi}, {Fumagalli}, {Leja}, {Lundgren},
  {Maseda}, {Nelson}, {Wake}, {Patel}, {Labb{\'e}}, {Faber}, {Grogin}, \&
  {Kocevski}}]{vanderWel2014}
{van der Wel}, A., {Franx}, M., {van Dokkum}, P.~G., {et~al.} 2014, \apj, 788,
  28

\bibitem[{{van der Wel} {et~al.}(2016){van der Wel}, {Noeske}, {Bezanson},
  {Pacifici}, {Gallazzi}, {Franx}, {Mu{\~n}oz-Mateos}, {Bell}, {Brammer},
  {Charlot}, {Chauk{\'e}}, {Labb{\'e}}, {Maseda}, {Muzzin}, {Rix}, {Sobral},
  {van de Sande}, {van Dokkum}, {Wild}, \& {Wolf}}]{vanderWel2016}
{van der Wel}, A., {Noeske}, K., {Bezanson}, R., {et~al.} 2016, \apjs, 223, 29

\bibitem[{{van Dokkum} {et~al.}(2008){van Dokkum}, {Franx}, {Kriek}, {Holden},
  {Illingworth}, {Magee}, {Bouwens}, {Marchesini}, {Quadri}, {Rudnick},
  {Taylor}, \& {Toft}}]{vanDokkum2008}
{van Dokkum}, P.~G., {Franx}, M., {Kriek}, M., {et~al.} 2008, \apjl, 677, L5

\bibitem[{{van Dokkum} {et~al.}(2010){van Dokkum}, {Whitaker}, {Brammer},
  {Franx}, {Kriek}, {Labb{\'e}}, {Marchesini}, {Quadri}, {Bezanson},
  {Illingworth}, {Muzzin}, {Rudnick}, {Tal}, \& {Wake}}]{vanDokkum2010}
{van Dokkum}, P.~G., {Whitaker}, K.~E., {Brammer}, G., {et~al.} 2010, \apj,
  709, 1018

\bibitem[{{Vergani} {et~al.}(2008){Vergani}, {Scodeggio}, {Pozzetti}, {Iovino},
  {Franzetti}, {Garilli}, {Zamorani}, {Maccagni}, {Lamareille}, {Le F{\`e}vre},
  {Charlot}, {Contini}, {Guzzo}, {Bottini}, {Le Brun}, {Picat}, {Scaramella},
  {Tresse}, {Vettolani}, {Zanichelli}, {Adami}, {Arnouts}, {Bardelli},
  {Bolzonella}, {Cappi}, {Ciliegi}, {Foucaud}, {Gavignaud}, {Ilbert},
  {McCracken}, {Marano}, {Marinoni}, {Mazure}, {Meneux}, {Merighi}, {Paltani},
  {Pell{\`o}}, {Pollo}, {Radovich}, {Zucca}, {Bondi}, {Bongiorno},
  {Brinchmann}, {Cucciati}, {de la Torre}, {Gregorini}, {Perez-Montero},
  {Mellier}, {Merluzzi}, \& {Temporin}}]{Vergani2008}
{Vergani}, D., {Scodeggio}, M., {Pozzetti}, L., {et~al.} 2008, \aap, 487, 89

\bibitem[{{Vettolani} {et~al.}(1997){Vettolani}, {Zucca}, {Zamorani}, {Cappi},
  {Merighi}, {Mignoli}, {Stirpe}, {MacGillivray}, {Collins}, {Balkowski},
  {Cayatte}, {Maurogordato}, {Proust}, {Chincarini}, {Guzzo}, {Maccagni},
  {Scaramella}, {Blanchard}, \& {Ramella}}]{Vettolani1997}
{Vettolani}, G., {Zucca}, E., {Zamorani}, G., {et~al.} 1997, \aap, 325, 954

\bibitem[{{Wake} {et~al.}(2012){Wake}, {van Dokkum}, \& {Franx}}]{Wake2012}
{Wake}, D.~A., {van Dokkum}, P.~G., \& {Franx}, M. 2012, \apjl, 751, L44

\bibitem[{{Williams} {et~al.}(2009){Williams}, {Quadri}, {Franx}, {van Dokkum},
  \& {Labb{\'e}}}]{Williams2009}
{Williams}, R.~J., {Quadri}, R.~F., {Franx}, M., {van Dokkum}, P., \&
  {Labb{\'e}}, I. 2009, \apj, 691, 1879

\bibitem[{{Wirth} {et~al.}(2004){Wirth}, {Willmer}, {Amico}, {Chaffee},
  {Goodrich}, {Kwok}, {Lyke}, {Mader}, {Tran}, {Barger}, {Cowie}, {Capak},
  {Coil}, {Cooper}, {Conrad}, {Davis}, {Faber}, {Hu}, {Koo}, {Le Mignant},
  {Newman}, \& {Songaila}}]{Wirth2004}
{Wirth}, G.~D., {Willmer}, C.~N.~A., {Amico}, P., {et~al.} 2004, \aj, 127, 3121

\bibitem[{{Woods} {et~al.}(2010){Woods}, {Geller}, {Kurtz}, {Westra},
  {Fabricant}, \& {Dell'Antonio}}]{Woods2010}
{Woods}, D.~F., {Geller}, M.~J., {Kurtz}, M.~J., {et~al.} 2010, \aj, 139, 1857

\bibitem[{{York} {et~al.}(2000){York}, {Adelman}, {Anderson}, {Anderson},
  {Annis}, {Bahcall}, {Bakken}, {Barkhouser}, {Bastian}, {Berman}, {Boroski},
  {Bracker}, {Briegel}, {Briggs}, {Brinkmann}, {Brunner}, {Burles}, {Carey},
  {Carr}, {Castander}, {Chen}, {Colestock}, {Connolly}, {Crocker}, {Csabai},
  {Czarapata}, {Davis}, {Doi}, {Dombeck}, {Eisenstein}, {Ellman}, {Elms},
  {Evans}, {Fan}, {Federwitz}, {Fiscelli}, {Friedman}, {Frieman}, {Fukugita},
  {Gillespie}, {Gunn}, {Gurbani}, {de Haas}, {Haldeman}, {Harris}, {Hayes},
  {Heckman}, {Hennessy}, {Hindsley}, {Holm}, {Holmgren}, {Huang}, {Hull},
  {Husby}, {Ichikawa}, {Ichikawa}, {Ivezi{\'c}}, {Kent}, {Kim}, {Kinney},
  {Klaene}, {Kleinman}, {Kleinman}, {Knapp}, {Korienek}, {Kron}, {Kunszt},
  {Lamb}, {Lee}, {Leger}, {Limmongkol}, {Lindenmeyer}, {Long}, {Loomis},
  {Loveday}, {Lucinio}, {Lupton}, {MacKinnon}, {Mannery}, {Mantsch}, {Margon},
  {McGehee}, {McKay}, {Meiksin}, {Merelli}, {Monet}, {Munn}, {Narayanan},
  {Nash}, {Neilsen}, {Neswold}, {Newberg}, {Nichol}, {Nicinski}, {Nonino},
  {Okada}, {Okamura}, {Ostriker}, {Owen}, {Pauls}, {Peoples}, {Peterson},
  {Petravick}, {Pier}, {Pope}, {Pordes}, {Prosapio}, {Rechenmacher}, {Quinn},
  {Richards}, {Richmond}, {Rivetta}, {Rockosi}, {Ruthmansdorfer}, {Sandford},
  {Schlegel}, {Schneider}, {Sekiguchi}, {Sergey}, {Shimasaku}, {Siegmund},
  {Smee}, {Smith}, {Snedden}, {Stone}, {Stoughton}, {Strauss}, {Stubbs},
  {SubbaRao}, {Szalay}, {Szapudi}, {Szokoly}, {Thakar}, {Tremonti}, {Tucker},
  {Uomoto}, {Vanden Berk}, {Vogeley}, {Waddell}, {Wang}, {Watanabe},
  {Weinberg}, {Yanny}, {Yasuda}, \& {SDSS Collaboration}}]{York2000}
{York}, D.~G., {Adelman}, J., {Anderson}, Jr., J.~E., {et~al.} 2000, \aj, 120,
  1579

\bibitem[{{Zahid} {et~al.}(2015){Zahid}, {Damjanov}, {Geller}, \&
  {Chilingarian}}]{Zahid2015}
{Zahid}, H.~J., {Damjanov}, I., {Geller}, M.~J., \& {Chilingarian}, I. 2015,
  \apj, 806, 122

\bibitem[{{Zahid} {et~al.}(2016{\natexlab{a}}){Zahid}, {Damjanov}, {Geller},
  {Hwang}, \& {Fabricant}}]{Zahid2016a}
{Zahid}, H.~J., {Damjanov}, I., {Geller}, M.~J., {Hwang}, H.~S., \&
  {Fabricant}, D.~G. 2016{\natexlab{a}}, \apj, 821, 101

\bibitem[{{Zahid} \& {Geller}(2017)}]{Zahid2017}
{Zahid}, H.~J., \& {Geller}, M.~J. 2017, \apj, 841, 32

\bibitem[{{Zahid} {et~al.}(2016{\natexlab{b}}){Zahid}, {Geller}, {Fabricant},
  \& {Hwang}}]{Zahid2016b}
{Zahid}, H.~J., {Geller}, M.~J., {Fabricant}, D.~G., \& {Hwang}, H.~S.
  2016{\natexlab{b}}, \apj, 832, 203

\end{thebibliography}

\end{document}